# Quantitative Law of Diffusion Induced Stress and Fracture

H.-J. Lei[1], H.-L. Wang[1,*], B. Liu[1,*], C.-A. Wang[2]

[1] *AML, CNMM, Department of Engineering Mechanics, Tsinghua University, Beijing, 100084, China*

[2] *State Key Lab of New Ceramics and Fine Processing, Department of Materials Science and Engineering, Tsinghua University, Beijing 100084, China*

∗ Corresponding authors, E-mail address: liubin@tsinghua.edu.cn (B. Liu); wang-hl06@mails.tsinghua.edu.cn (H.-L. Wang)

**Abstract:**
   In diffusion processes of solid materials, such as in the classical thermal shock problem and the recent lithium ion battery, the maximum diffusion induced stress (DIS) is a very important quantity. However a widely accepted, accurate and easy-to-use quantitative formula on it still lacks. In this paper, by normalizing the governing equations, an almost analytical model is developed, except a single-variable function of the dimensionless Biot number which cannot be determined analytically and is then given by a curve. Formulae for various typical geometries and working conditions are presented. If the stress and the diffusion process are fully coupled (i.e. stress-dependent diffusion), as in lithium ion diffusion, the normalized maximum DIS can be characterized by a two-variable function of a dimensionless coupling parameter and the Biot number, which is obtained numerically and presented in contour plots. Moreover, it is interesting to note that these two parameters, within a wide range, can be further approximately combined into a single dimensionless parameter to characterize the maximum DIS. These formulae together with curves/contours provide engineers and materialists a simple and easy way to quickly obtain the stress and verify the reliability of materials under various typical diffusion conditions. Via energy balance analysis, the model of diffusion induced fracture is also developed. It interestingly predicts that the spacing of diffusion induced cracks is constant, independent of the thickness of specimen and the concentration difference. Our thermal shock experiments on alumina plates validate these qualitative and quantitative theoretical predictions, such as the constant crack spacing and the predicted critical temperature difference at which the cracks initiate. Furthermore, the proposed model can interpret the observed hierarchical crack patterns for high temperature jump cases. The implication of our study to practical designing is that a specimen with smaller thickness or radius can sustain more dramatic diffusion processes safely, and if its dimension perpendicular to the diffusion direction is smaller than the predicted crack spacing, no diffusion can lead to any fracture. We also suggest an easy way, by using the proposed concise relation, to determine the fracture toughness by simply measuring the strength and the thermal shock induced crack-spacing.

## 1. Introduction

Diffusion and diffusion induced fracture are widely observed phenomena in nature and industry. Diffusion process is the directional migration of particles or energy driven by a physical or chemical gradient, such as the temperature gradient, concentration gradient, chemical potential gradient, and so on. Diffusion process often causes a volume change in solids. Due to the non-uniform distribution of the diffusion species, the volume change is usually inhomogeneous and results in stresses, i.e. diffusion induced stresses (DISes). Once the DIS exceeds the strength of the material, fracture happens.

Diffusion induced fracture is the major cause of the deterioration and failure of some materials, devices and structures. Therefore, it is a hot topic for scientific research. For particles diffusion, most recent research attentions are focused on the lithiation in the lithium battery, which causes large volumetric expansion (even up to 400%) of the silicon electrode (Qi and Harris, 2010). The so-called electrochemical shock fracture (Woodford et al., 2012) has been observed in experiments with the help of X-ray diffraction (XRD) (Thackeray et al., 1998), scanning electron microscopy (SEM) (Lim et al., 2001), tunneling electron microscopy (TEM) (Thackeray et al., 1998), NMR spectroscopy (Tucker et al., 2002). Several scholars have studied lithium diffusion, emphasizing on the analysis of DISes and fractures (Prussin,1961; Li,1978; Huggins and Nix, 2000; Yang,2005; Christensen and Newman, 2006; Verbrugge and Cheng, 2009; Cheng and Verbrugge, 2010a, b; Woodford et al., 2010; Deshpande et al., 2011) . In particular, Huggins and Nix (2000) developed a simple bilayer plate model to describe fracture associated with decrepitation during battery cycles. Without considering the effects of concentration gradient, a mathematical model for diffusion-induced fracture has been developed by Christensen and Newman (2006). Cheng and Verbrugge (2010a, b) derived the DIS solutions in infinite series forms, and proposed an approximate analytical model for studying fracture in electrodes. In order to obtain more precise results, many numerical simulations have been carried out (Zhang et al., 2007; Bhandakkar and Gao, 2010; Park et al., 2011; Shi et al., 2011; Purkayastha and McMeeking, 2012; Bower and Guduru, 2012; Zhao et al., 2012). For example, Bhandakkar and Gao (2010) developed a cohesive model on crack nucleation in a strip electrode during galvanostatic intercalation and deintercalation processes, and a critical characteristic dimension is identified, below which crack nucleation becomes impossible. Bower and Guduru (2012) proposed a simple mixed finite element method in which the governing equations for diffusion and equilibrium are fully coupled. Based on first-principles calculations of the atomic-scale structural and electronic properties in a model amorphous silicon (a-Si) structure, Zhao et al., (2012) provided a detailed picture of the origin of changes in the mechanical properties. Besides the extensive studies on the failure due to DIS, some researchers have explored the ways to improve the properties of lithium batteries by optimizing the shape and dimension of the electrode and its constituent particles (Zhang et al., 2007; Park et al., 2011; Ryu et al., 2011; Xiao et al., 2011; Vanimisetti and Ramakrishnan, 2012; Lim et al., 2012). Zhang et al. (2007) developed a three-dimensional finite element model of spherical and ellipsoidal shape particles to simulate DISes. They claimed that ellipsoidal particles with large aspect ratios are preferred to reduce the intercalation-induced stresses. Ryu et al. (2011) used a unique transmission electron microscope (TEM) technique to show that Si nanowires (NWs) with diameters in the range of a few hundred nanometers can be fully lithiated and delithiated without fracture. Considering the critical size for the crack gap in continuous films, Xiao et al. (2011) introduced a simple patterning approach to improve the cycling stability of silicon electrode.

For energy diffusion, the fracture induced by heat diffusion or the so-called thermal shock is one of the most important problems. The thermal shock resistance, as one of characteristic parameters, is generally measured by a critical temperature difference, at which the strength of brittle materials catastrophically decreases (Bahr et al., 1986; Swain, 1990; Lu and Fleck, 1998; Liu et al., 2009). In order to predict the critical temperature difference, Manson proposed a semi-empirical solution, but the solution was inaccurate when the Biot number $Bi > 20$ (Manson, 1954). Hasselman determined the critical temperature difference using the minimum energy method with assumptions that materials are entirely brittle and contain uniformly distributed circular micro-cracks (Hasselma.Ph, 1969). The crack patterns including crack densities and morphologies after thermal shock are also important information in understanding thermal shock resistance of materials, which have been studied in Ref. (Erdogan and Ozturk, 1995; Hutchinson and Xia, 2000; Collin and Rowcliffe, 2002; Bohn et al., 2005). Bahr et al. carried out several important studies on the scaling behavior of parallel crack patterns driven by thermal shock (Bahr et al., 1986; Bahr et al., 2010). Moreover, the thermal shock resistance of ceramics used in thermal protection system has been extensively studied (Levine et al., 2002; Fahrenholtz et al., 2007; Ma and Han, 2010; Monteverde et al., 2010). Inspired by biological microstructure, Song et al. (2010) have presented a novel surface-treatment method to improve the thermal shock resistance of materials.

Although there are many analytical models on electrochemical shock or thermal shock, an analytical model that is simple, accurate and easy use for materialists and engineers is still lacked. In this paper, a concise diffusion-induced-failure model based on normalized governing equations and energy balance principle is developed. The structure of this paper is as follows. From Section 2 to Section 5, we focus on the contraction diffusion (e.g., quenching) induced stress and fracture. In Section 2, we first analytically determine the maximum DIS and the crack spacing of an infinite plate. The results are then validated by our thermal shock experiments. The effects of the mechanical/diffusion boundaries and geometry are discussed in Section 3 and Section 4, respectively. A model accounting for fully coupling between the diffusion and stress is presented in Section 5. Results of the expansion diffusion (e.g., fast heating-up) induced maximum stresses are presented in Section 6. Conclusions are summarized in Section 7.

## 2. Analysis on diffusion induced failure of a free infinite plate

2.1 Diffusion variable and its governing equation

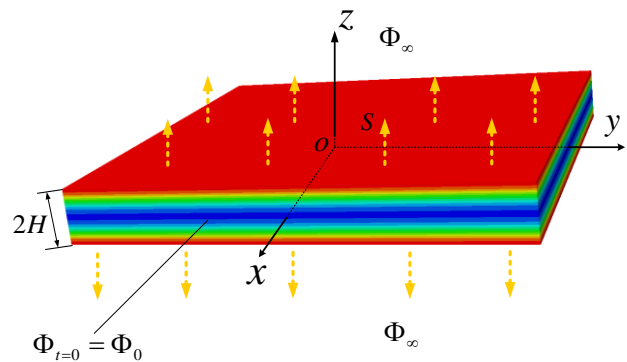

**Fig.1.** The schematic diagram of diffusion on an infinite plate. The top and bottom surfaces of the plate are traction free. The red color represents tensile stress and blue color means compressive stress.

All diffusions obey the similar principle. To expand the applicability of our model, we introduce a general diffusion variable $\Phi$. For energy diffusion, such as heat conducting, this diffusion variable is the temperature. For particles diffusion, the diffusion variable is the concentration of the diffusing species. Usually, the flux of diffusion variable $\mathbf{J}$ is assumed to be proportional to the gradient of $\Phi$, i.e.

$$\mathbf{J} = -D\nabla\Phi \tag{1}$$

where $D$ is the diffusion coefficient (or the thermal conductivity). Conservation of the diffusion variable yields

$$\frac{\partial \Phi}{\partial t} + \nabla \cdot \mathbf{J} = 0 \tag{2}$$

Substituting Eq. (1) into Eq. (2), the governing equation of diffusion is

$$\frac{\partial \Phi}{\partial t} = D\nabla^2\Phi \tag{3}$$

To determine the diffusion variable, the boundary conditions and initial conditions are also needed. Since diffusions in plates are widely observed in practical applications, such as the electrochemical shock in flat battery or the thermal shock in thermal barrier coating, we first investigate an infinite plate with a thickness $2H$ immersed in an environment where the diffusion variable is $\Phi_\infty$, as shown in Fig.1. Obviously, the diffusion variable only varies along *z* direction. Assuming that the diffusion variable of the plate is initially uniform with the value of $\Phi_0$, the initial condition is

$$\Phi\big|_{t=0} = \Phi_0 \tag{4}$$

Due to the symmetry, the boundary conditions for the upper half part of the plate (see the inset of Fig.2) are

$$D\frac{\partial \Phi}{\partial z}\bigg|_{z=H} = -S(\Phi\big|_{z=H} - \Phi_\infty) \tag{5}$$

$$\frac{\partial \Phi}{\partial z}\bigg|_{z=0} = 0 \tag{6}$$

where $S$ is the interface diffusion coefficient. For the diffusion of lithium ion,

$$S = \frac{(k'_a + k'_c)}{c} \tag{7}$$

where $k'_a$ and $k'_c$ are the interfacial reaction-rate constant, $c$ is the concentration of the total site available for insertion within the host particle (Cheng and Verbrugge, 2010a, b). For the diffusion of heat,

$$S = \frac{h}{\rho c_p} \tag{8}$$

where $h$ is the surface heat exchange coefficient, $\rho$ is the mass density and $c_p$ is the heat capacity.

Equations (3)-(6) can be normalized using the following normalized variables

$$\hat{\Phi} = \frac{\Phi - \Phi_0}{\Phi_\infty - \Phi_0}$$

$$\hat{t} = \frac{Dt}{H^2}$$

$$\hat{z} = \frac{z}{H}$$

$$Bi = \frac{SH}{D}$$

(9a-d)

where $\hat{\Phi}$ is the normalized diffusion variable, $\hat{t}$ is the normalized time, $\hat{z}$ is the normalized coordinate, $Bi$ is the Biot number represents the normalized ratio between the diffusion capacities over the interface and in the bulk material.

The governing equation becomes

$$\frac{\partial \hat{\Phi}(\hat{z},\hat{t})}{\partial \hat{t}} = \frac{\partial^2 \hat{\Phi}(\hat{z},\hat{t})}{\partial \hat{z}^2} \tag{10}$$

and the corresponding boundary/initial conditions are

$$\left.\frac{\partial \hat{\Phi}}{\partial \hat{z}}\right|_{\hat{z}=1} = -Bi(\hat{\Phi}|_{\hat{z}=1} - 1) \tag{11}$$

$$\left.\frac{\partial \hat{\Phi}}{\partial \hat{z}}\right|_{\hat{z}=0} = 0 \tag{12}$$

$$\hat{\Phi}|_{\hat{t}=0} = 0 \tag{13}$$

2.2 The contraction diffusion-induced maximum stress of a plate

Once the field of diffusion variable is known, the DIS can be obtained. The constitutive equations for linear elastic material are

$$\sigma_{11} = \frac{E}{(1+v)(1-2v)}\left[(1-2v)\varepsilon_{11} + v(\varepsilon_{11} + \varepsilon_{22} + \varepsilon_{33})\right] - \frac{E}{(1-2v)}\varepsilon_\Phi$$

$$\sigma_{22} = \frac{E}{(1+v)(1-2v)}\left[(1-2v)\varepsilon_{22} + v(\varepsilon_{11} + \varepsilon_{22} + \varepsilon_{33})\right] - \frac{E}{(1-2v)}\varepsilon_\Phi \tag{14}$$

$$\sigma_{33} = \frac{E}{(1+v)(1-2v)}\left[(1-2v)\varepsilon_{33} + v(\varepsilon_{11} + \varepsilon_{22} + \varepsilon_{33})\right] - \frac{E}{(1-2v)}\varepsilon_\Phi$$

where $\sigma_{11}, \sigma_{22}, \sigma_{33}$ and $\varepsilon_{11}, \varepsilon_{22}, \varepsilon_{33}$ are stresses and strains along 1-, 2-, 3- coordinate directions respectively, $E$ is the Young's modulus, $v$ is the Poisson's ratio. $\varepsilon_\Phi$ is the diffusion-induced strain which is

$$\varepsilon_\Phi = \alpha \Delta \Phi \tag{15}$$

where $\alpha$ is the coefficient of diffusion-induced expansion and $\Delta \Phi = \Phi - \Phi_0$.

In this plane case, the constitutive equation is

$$\sigma_{xx} = \frac{E}{1-v^2}(\varepsilon_{xx} + v\varepsilon_{yy}) - \frac{E}{1-v}\varepsilon_\Phi$$

$$\sigma_{yy} = \frac{E}{1-v^2}(\varepsilon_{yy} + v\varepsilon_{xx}) - \frac{E}{1-v}\varepsilon_\Phi \tag{16}$$

$$\sigma_{zz} = 0$$

in the Cartesian coordinate system, where $x$-, $y$-, $z$- correspond to 1-, 2-, 3- in Eq. (14)

respectively. Because all the boundaries of the plate are traction free, we have the self-balance condition

$$\int_0^H \sigma_{xx}(z,t)dz = 0 \tag{17}$$

The in-plane strains $\varepsilon_{xx} = \varepsilon_{yy}$ can be solved by substituting Eq.(15) and Eq.(16) into Eq.(17), and the stress becomes

$$\begin{aligned}\sigma_{xx}(z,t) &= \frac{E\alpha}{(1-v)}\left[\frac{1}{H}\int_0^H \Delta\Phi(z,t)dz - \Delta\Phi(z,t)\right] \\ &= \frac{E\alpha}{(1-v)}\left[\int_0^1 \hat{\Phi}(\hat{z},\hat{t},Bi)\Delta\Phi_\infty d\hat{z} - \hat{\Phi}(\hat{z},\hat{t},Bi)\Delta\Phi_\infty\right] \\ &= -\frac{E\alpha\Delta\Phi_\infty}{(1-v)}\hat{g}_I(\hat{z},\hat{t},Bi)\end{aligned} \tag{18}$$

where

$$\hat{g}_I(\hat{z},\hat{t},Bi) = \hat{\Phi}(\hat{z},\hat{t},Bi) - \int_0^1 \hat{\Phi}(\hat{z},\hat{t},Bi)d\hat{z} \tag{19}$$

is the normalized DIS and $\Delta\Phi_\infty = \Phi_\infty - \Phi_0$.

We assume that fracture can only be caused by tensile stress. Therefore during the contraction diffusion process (i.e., generalized "quenching") where $\alpha\Delta\Phi_\infty$ is negative, the maximum $\hat{g}_I(\hat{z},\hat{t},Bi)$ corresponds to the maximum DIS, while during the expansion diffusion process (i.e., generalized "heating-up") the minimum of $\hat{g}_I(\hat{z},\hat{t},Bi)$ corresponds to the maximum DIS. In this section we focus on the generalized "quenching" process, while the generalized "heating-up" case is studied in Section 6. Therefore, we need to obtain the maximum $\hat{g}_I(\hat{z},\hat{t},Bi)$ which obviously emerges on the surface $z = H$, i.e., $\hat{z} = 1$ (see the inset in Fig.2). A critical time $\hat{t}_{max}$ exists at which the DIS is largest, because the DIS is both zero at the very beginning of diffusion and after sufficiently long time when the diffusion variable in the plate is uniform. The critical time $\hat{t}_{max}$ is determined by

$$\left.\frac{\partial \hat{g}_I(1,\hat{t},Bi)}{\partial \hat{t}}\right|_{\hat{t}=\hat{t}_{max}} = 0 \tag{20}$$

We can see that $\hat{t}_{max}$ only depends on the Biot number $Bi$, and the maximum normalized DIS

$$\hat{g}_I[1,\hat{t}_{max}(Bi),Bi] \equiv g_I^{max}(Bi) \tag{21}$$

is also determined by the Biot number $Bi$ only. The subscript "I" of $g_I^{max}(Bi)$ and $\hat{g}_I$ represents the mode. In this paper, different loading, boundary and geometry are categorized as Mode I to Mode VII. For example "Mode I" represents the diffusion in a plate with free in-plane expansion and insulated bottom boundary. So the contraction diffusion induced maximum stress $\sigma^{max}$ can be written as

$$\sigma^{max} = -\frac{E\alpha\Delta\Phi_\infty}{1-v}g_I^{max}(Bi) \tag{22}$$

It should be emphasized that the maximum DIS $\sigma^{max}$ has then been expressed in a simple analytical formula except a single variable function $g_I^{max}(Bi)$, which can be given by a curve obtained numerically as follows. Therefore, it will be convenient for researchers and engineers to obtain the maximum DIS $\sigma^{max}$ quickly.

There are several ways to solve Eqs. (10)-(13), such as finite difference method. In order to obtain the accurate solutions quickly, here we adopt an analytical solution in infinite series form by the separation of variables method, which gives rises to

$$\hat{\Phi} = -\sum_{n=1}^{\infty} \frac{4\sin\omega_n \cos(\omega_n \hat{z})}{2\omega_n + \sin(2\omega_n)} e^{-\omega_n^2 \hat{t}} + 1 \tag{23}$$

where $\omega_n$ is the n-th positive root of the following equation

$$\omega \tan \omega = Bi \tag{24}$$

The procedure to obtain (23) can be found in most textbooks regarding the solution of partial differential equations and is presented in Appendix A. Substituting Eq. (23) into Eq. (19) to obtain $\hat{g}_I$ and maximize it numerically, we plot $g_I^{max}(Bi)$, *i.e.* the normalized maximum DIS, as a function of Biot number $Bi$, as shown in Fig 2. $g_I^{max}(Bi)$ is a monotonically increasing function and approaches 1 when $Bi$ tends to infinity. For more convenient calculation, an approximate fitting function of $g_I^{max}(Bi)$ is also given as,

$$g_I^{max}(Bi) = \begin{cases} 0.3 - 0.3e^{-Bi} & (0.001 \le Bi < 0.1) \\ 0.652 - 0.178e^{-Bi} - 0.466e^{-\frac{Bi}{5}} & (0.1 \le Bi \le 10) \\ 0.88 - 0.262e^{-\frac{Bi}{10}} - 0.228e^{-\frac{Bi}{50}} & (10 < Bi \le 100) \end{cases} \tag{25}$$

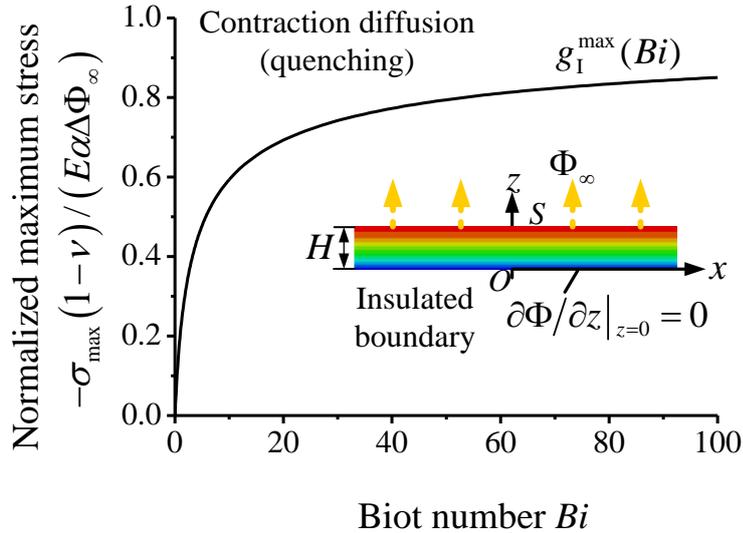

**Fig.2.** The normalized maximum diffusion induced stress varies with Biot number $Bi$ for the case of diffusion in a plate with free in-plane expansion and insulated boundary. The inset contour plot is an illustration of the stress distribution in the material. The red color represents tensile stress and blue color means compressive stress.

## 2.3 The diffusion-induced failure analysis

Once the maximum DIS is obtained, we perform failure analysis to determine the critical difference of the diffusion variable $\Delta\Phi_\infty = \Phi_\infty - \Phi_0$ that leads to fracture, and the corresponding crack density. If the maximum DIS is lower than the failure strength $\sigma_f$, no fracture happens. Otherwise, cracks emerge. The critical $\Delta\Phi_\infty$ corresponds to $\sigma^{max} = \sigma_f$ and can be determined from Eq. (22) as

$$\Delta\Phi_{cr} = -\frac{\sigma_f(1-v)}{E\alpha g_I^{max}(Bi)} \tag{26}$$

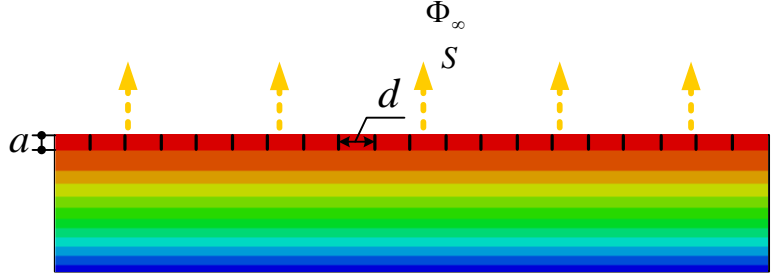

**Fig.3.** The schematic diagram for the failure analysis of single-level crack pattern. The red color represents tensile stress and blue color means compressive stress.

The average crack spacing $d$ can be estimated by the energy balance of the top layer of the plates. As shown in Fig.3, it is assumed that the elastic strain energy of the fractured region is completely converted into the surface energy of cracks. So we obtain $2\gamma a = \dfrac{\sigma_f^2 a d}{2E}$, where $\gamma$ is the surface energy per unit area, $a$ is the crack length at the very beginning. Considering the fracture toughness $K_{IC} = \sqrt{2\gamma E}$, the crack spacing can be summarized as following

$$d = \begin{cases} \infty \text{ (no crack)}, & |\Delta\Phi| < |\Delta\Phi_{cr}| \\ \dfrac{2K_{IC}^2}{\sigma_f^2}, & |\Delta\Phi| \geq |\Delta\Phi_{cr}| \end{cases} \tag{27}$$

It is interesting to note that the crack-spacing $d$ only depends on the mechanical properties, independent of diffusion loading and conditions. One can imagine that if the size of platelet is smaller than $d$, no fracture will happen anymore. A biomimetic staggered brick-mortar-like microstructure can then be adopted to construct platelet reinforced composites subject to diffusion (Lei et al., 2012; Xiao et al., 2011).

Equation (27) also provides an easy way to determine the fracture toughness $K_{IC}$ by simply measuring the thermal shock induced crack-spacing $d$ and the strength $\sigma_f$.

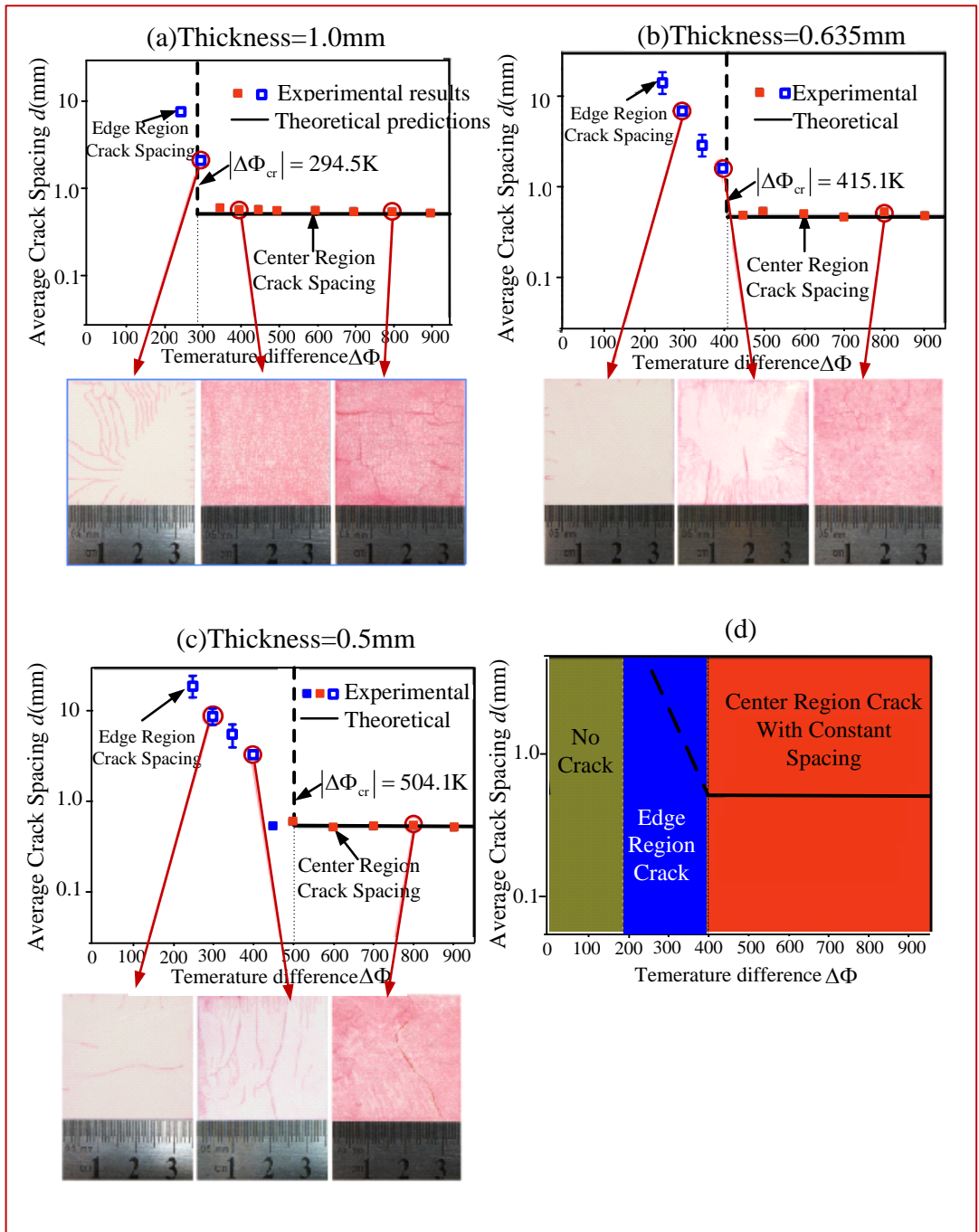

**Fig.4.** Average crack spacing and patterns of specimens with different thickness (a) 1mm, (b) 0.635mm and (c) 0.5mm, subject to different temperature difference. (d) The schematic map of crack spacing distribution with temperature difference.

2.4 Experimental validation and further discussion

We carry out a series of thermal shock experiments on alumina plates to validate our model. In this heat diffusion-induced failure, $\Delta\Phi$ and $\alpha$ are the temperature difference and the coefficient of thermal expansion, respectively. The specimens are commercial products made of particles whose radius are $3\pm 1\mu m$ by casting method with relative density of 99%. The Young's modulus $E=386 GPa$, Poisson's ratio $v=0.254$, the fracture toughness $K_{IC}=4 MPa.m^{1/2}$, the failure strength $\sigma_f=250MPa$, the coefficient of thermal expansion $\alpha=8.52\times 10^{-6} K^{-1}$, the surface heat exchange coefficient $h = 25000 W/(m^2 \cdot K)$ and the thermal conductivity $\lambda = 14 W/(m \cdot K)$. The geometries of the plates are squares with side length of 35mm and with three different thicknesses, i.e. 1.0mm, 0.65mm and 0.5mm. The thermal shock test system is composed of a furnace and a water tank. Specimens are first heated at the furnace for an hour, then dropped into water tank vertically through a pipe. The temperature difference $|\Delta\Phi|$ for all specimens are chosen as 100K, 200K, 250K, 300K, 350K, 400K, 450K, 500K, 600K, 700K, 800K and 900K. After thermal shock, all specimens are dyed by red ink for easy observation of crack patterns.

The average crack spacing as a function of the temperature difference for different plate thicknesses is shown in Fig.4(a-c), and each point represents the average of three repeating experiments. Typical crack patterns in experiments are also given in the figure. We can find that there is a critical temperature difference $\Delta\Phi_{cr}$ just as predicted by our theoretical model Eq. (26) at which the uniform crack patterns start emerging at the central region.

Moreover, it is very interesting to note that beyond this critical temperature, the crack spacing at the central region always keeps constant, i.e. $d \approx 0.51$mm, independent of the temperature difference and the thickness of the specimen, which is also in good agreement with our theoretical prediction Eq. (27). Since $g_I^{max}(Bi)$ is a monotonically increasing function and $Bi = SH/D$, the thicker plate corresponds to the larger thermal stress or the smaller critical temperature difference $\Delta\Phi_{cr}$. It should be pointed out that our theoretical model is developed for infinite large plates, and it can predict central region successfully (shown by the red solid squares in Fig. 4). The emergency of cracks in the edge region (shown by open blue squares in Fig. 4) depends on more complex factors, such as initial edge defects, three-dimensional heat transfer and resulting thermal stress, which are beyond the scope of our current theoretical model.

Fig 5(a) and (b) show the crack patterns of alumina after thermal shock at the temperature difference $|\Delta\Phi| = 400K$ and $|\Delta\Phi| = 900K$, and we can find a hierarchical crack pattern in the latter situation. It is interesting to note that there are many similar hierarchical-crack phenomena in the nature and engineering, such as earth cracks in drought days and cracking of electrodes in lithium batteries as shown in Fig.5(c) and (d), although with significantly different dimension. They are all characterized by the combination of the diffusion process and the fracture process to reduce the strain energy, and can also be interpreted by our model. The first level crack emerges and propagates when the thermal stress equals to the failure strength. After that, if the temperature gradient is still high enough at the first-level crack tips, some of first-level cracks will propagate again, and hierarchical crack patterns hence form. We can still adopt previous energy balance analysis while noting that the failure strength at this level is much lower than $\sigma_f$ due to the existence of the first-level cracks. According to Eq. (27), the spacing of the second-level crack is therefore much larger than that for the first level.

The relation between the strength of plate and temperature difference is schematically shown in Fig.6. When the temperature difference $\Delta\Phi = \Delta\Phi_{cr}^{(1)}$, the first-level crack is initiated, so the strength is reduced from $\sigma_f$ to $\sigma_c$ suddenly. The strength will drop again as the emerging of the second-level crack. i.e., if $\Delta\Phi = \Delta\Phi_{cr}^{(2)}$.

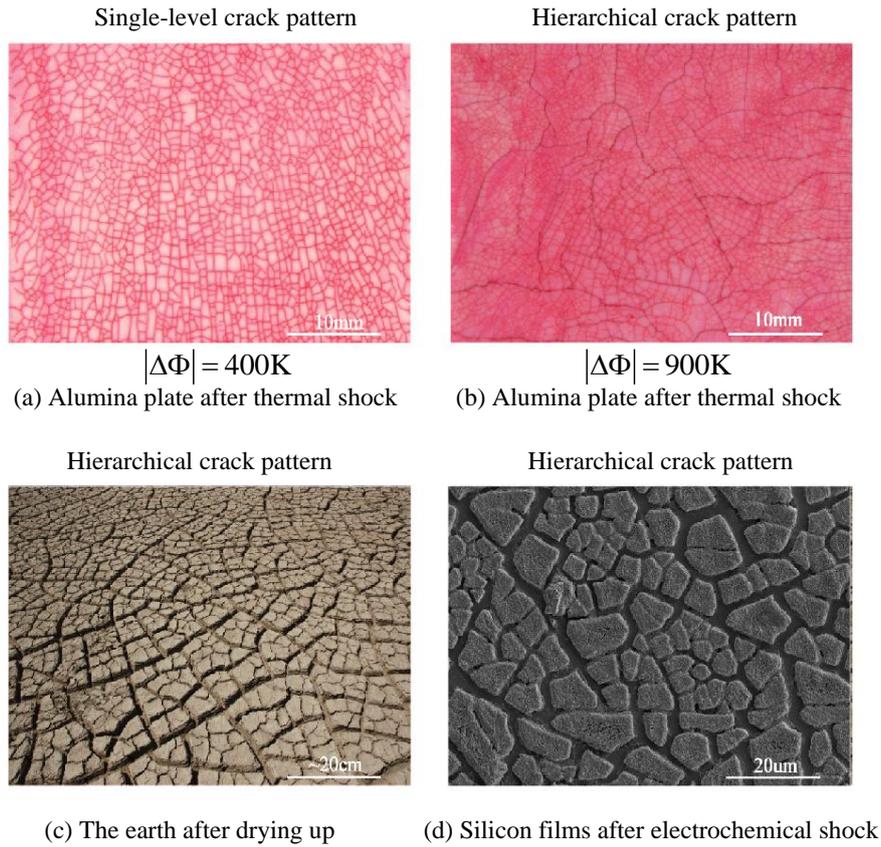

**Fig.5.** The picture of alumina plate with thickness of 1 mm after thermal shock at (a) temperature difference of 400K.; (b) temperature difference of 900K.(c) The image of dried-earth cracks. (d) An SEM image showing surface morphology of 500 nm thick Si films after ten electrochemical cycles in lithium battery(Li et al., 2011) (Reproduced by permission of ECS—The Electrochemical Society).

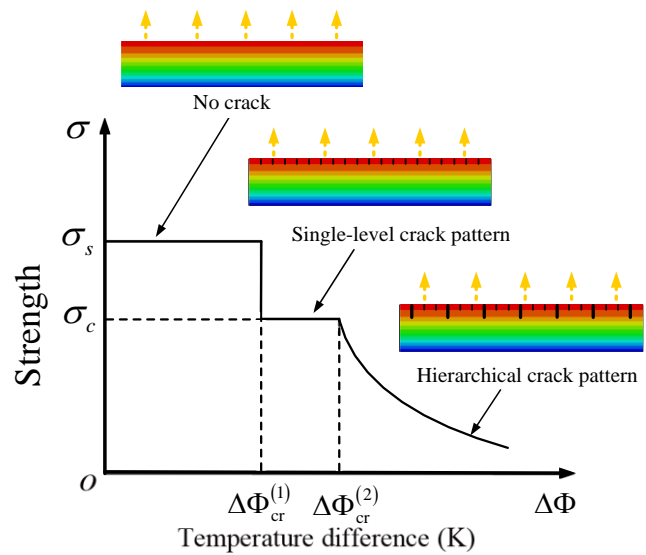

**Fig.6.** The schematic figure for variation of strength as a function of temperature difference. The red color represents tensile stress and blue color means compressive stress.

## 3. The maximum diffusion induced stress for different boundary conditions

Section 2 presents the diffusion in a plate with free in-plane expansion and insulated boundary, and similar derivations can be carried out for other cases with different mechanical and diffusion boundary conditions. In the following, we will study other three typical and extreme cases which are named as mode II-IV, and the case in Section 2 is named as mode I.

*Mode II Contract diffusion in a plate with free in-plane expansion and fixed diffusion variable at the bottom*

In this case (see the inset of Fig.7 (a)), the boundary conditions are $D\partial\Phi/\partial z|_{z=H} = -S(\Phi|_{z=H} - \Phi_\infty)$ for the upper surface and $\Phi|_{z=0} = \Phi_0$ for the bottom surface. Correspondingly, the normalized diffusion induced maximum stress $g_{II}^{max}(Bi)$ is shown by the stars in Fig.7 (a)(b). $g_{II}^{max}(Bi)$ can also be approximately fitted in a similar way to Eq. (25), which is summarized in Table 1 together with other cases.

*Mode III Contraction diffusion in a plate with rigid in-plane constraint and insulated bottom boundary*

The boundary conditions in this case are $D\partial\Phi/\partial z|_{z=H} = -S(\Phi|_{z=H} - \Phi_\infty)$ and $\partial\Phi/\partial z|_{z=0} = 0$ as shown in the inset of Fig. 7 (a). The rigid in-plane constraint implies $\varepsilon_{xx} = \varepsilon_{yy} = 0$. Using Eq. (16), the stress becomes

$$\sigma_{xx}(z,t) = -\frac{E\alpha\Delta\Phi(z,t)}{1-v} \quad (28)$$

The maximum DIS therefore corresponds to the maximum magnitude of $\Delta\Phi(z,t)$. As time goes by, the temperature of the whole plate will approach $\Phi_\infty$ finally, so

$$\frac{\sigma^{max}}{-E\alpha\Delta\Phi_\infty/(1-v)} = g_{III}^{max}(Bi) = 1 \quad (29)$$

as shown by the horizontal solid line in Fig.7(a).

*Mode IV Contraction diffusion in a plate with rigid in-plane constraint and fixed diffusion variable at the bottom*

The boundary conditions in this case (see the inset of Fig.7 (a)) are $D\partial\Phi/\partial z|_{z=H} = -S(\Phi|_{z=H} - \Phi_\infty)$ and Eq. (28) still holds due to the rigid in-plane constraint $\varepsilon_{xx} = \varepsilon_{yy} = 0$. The maximum magnitude of $\Delta\Phi(z,t)$ occurs on the upper surface ($\hat{z}=1$) after sufficiently long time. The diffusion field finally evolves into a steady state and the upper surface diffusion variable becomes $\left[\Phi_0 + (\Phi_\infty - \Phi_0)Bi/(1+Bi)\right]$. The normalized maximum DIS $g_{IV}^{max}(Bi)$ is shown by the curved solid line in Fig.7 (a) and expressed as

$$\frac{\sigma^{max}}{-E\alpha\Delta\Phi_\infty/(1-v)} = g_{IV}^{max}(Bi) = \frac{Bi}{1+Bi} \quad (30)$$

**Table 1.** The approximate fitting functions of the maximum diffusion induced stresses in stress-independent diffusion

| | | |
|---|---|---|
| Plate | 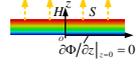 Case I<br>Free in-plane expansion, insulated bottom boundary | |
| | $g_{\text{I}}^{\max}(Bi) = \begin{cases} 0.3 - 0.3e^{-Bi} & (0.001 \leq Bi < 0.1) \\ 0.652 - 0.178e^{-Bi} - 0.466e^{-\frac{Bi}{5}} & (0.1 \leq Bi \leq 10) \\ 0.88 - 0.262e^{-\frac{Bi}{10}} - 0.228e^{-\frac{Bi}{50}} & (10 < Bi \leq 100) \end{cases}$ | $-g_{\text{I}}^{\min}(Bi) = \begin{cases} 0.162 - 0.162e^{-Bi} & (0.001 \leq Bi < 0.1) \\ 0.279 - 0.112e^{-Bi} - 0.166e^{-\frac{Bi}{5}} & (0.1 \leq Bi \leq 10) \\ 0.306 - 0.0797e^{-\frac{Bi}{10}} - 0.0225e^{-\frac{Bi}{50}} & (10 < Bi \leq 100) \end{cases}$ |
| | 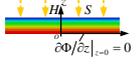 Case II<br>Free in-plane expansion, fixed diffusion variable at the bottom | |
| | $g_{\text{II}}^{\max}(Bi) = \begin{cases} 0.31 - 0.31e^{-Bi} & (0.001 \leq Bi < 0.1) \\ 0.649 - 0.19e^{-Bi} - 0.45e^{-\frac{Bi}{5}} & (0.1 \leq Bi \leq 10) \\ 0.88 - 0.262e^{-\frac{Bi}{10}} - 0.228e^{-\frac{Bi}{50}} & (10 < Bi \leq 100) \end{cases}$ | $-g_{\text{II}}^{\min}(Bi) = \begin{cases} 0.478 - 0.477e^{-Bi} & (0.001 \leq Bi < 0.1) \\ 0.474 - 0.293e^{-Bi} - 0.149e^{-\frac{Bi}{5}} & (0.1 \leq Bi \leq 10) \\ 0.497 - 0.0711e^{-\frac{Bi}{10}} - 0.0180e^{-\frac{Bi}{50}} & (10 < Bi \leq 100) \end{cases}$ |
| Cylinder | 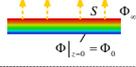 Case V<br>Fixed axial constraint | |
| | $g_{\text{V},\theta}^{\max}(Bi) = \begin{cases} 0.239 - 0.239e^{-Bi} & (0.001 \leq Bi < 0.1) \\ 0.56 - 0.0975e^{-Bi} - 0.45e^{-\frac{Bi}{5}} & (0.1 \leq Bi \leq 10) \\ 0.837 - 0.26e^{-\frac{Bi}{10}} - 0.284e^{-\frac{Bi}{50}} & (10 < Bi \leq 100) \end{cases}$ | $-g_{\text{V},r}^{\min}(Bi) = -g_{\text{V},\theta}^{\min}(Bi)$<br>$= \begin{cases} 0.121 - 0.121e^{-Bi} & (0.001 \leq Bi < 0.1) \\ 0.213 - 0.0824e^{-Bi} - 0.130e^{-\frac{Bi}{5}} & (0.1 \leq Bi \leq 10) \\ 0.233 - 0.0621e^{-\frac{Bi}{10}} - 0.0169e^{-\frac{Bi}{50}} & (10 < Bi \leq 100) \end{cases}$ |
| | $g_{\text{V},z}^{\max}(Bi, \nu) = \max(1 - \nu, f(Bi, \nu))$ $(0.01 \leq \nu < 0.49)$<br>$f = \begin{cases} \frac{1}{(1+0.842\nu)}\left(0.612 - 0.612e^{-Bi}\right) & (0.001 \leq Bi < 0.1) \\ \frac{1}{(1+0.982\nu)}\left(1 - 0.581e^{-Bi} - 0.437e^{-\frac{Bi}{5}}\right) & (0.1 \leq Bi \leq 10) \\ \frac{1}{(1+0.431\nu)}\left(1.02 - 0.141e^{-\frac{Bi}{10}} - 0.149e^{-\frac{Bi}{50}}\right) & (10 < Bi \leq 100) \end{cases}$ | $-g_{\text{V},z}^{\min}(Bi, \nu)$ $(0.01 \leq \nu < 0.49)$<br>$= \begin{cases} \frac{1}{(1-1.67\nu)}\left(0.0156 - 0.0156e^{-Bi}\right) & (0.001 \leq Bi < 0.1) \\ \frac{1}{(1-1.62\nu)}\left(0.0451 - 0.0113e^{-Bi} - 0.0345e^{-\frac{Bi}{5}}\right) & (0.1 \leq Bi \leq 10) \\ \frac{1}{(1-1.60\nu)}\left(0.0525 - 0.0181e^{-\frac{Bi}{10}} - 0.0052e^{-\frac{Bi}{50}}\right) & (10 < Bi \leq 100) \end{cases}$ |
| | 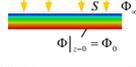 Case VI<br>Free axial constraint | |
| | $g_{\text{VI},\theta}^{\max}(Bi) = g_{\text{VI},z}^{\max}(Bi) = g_{\text{VI},\theta}^{\max}(Bi)$ | $-g_{\text{VI},r}^{\min}(Bi) = -g_{\text{VI},\theta}^{\min}(Bi) = -g_{\text{V},r}^{\min}(Bi) = -g_{\text{V},\theta}^{\min}(Bi)$<br>$-g_{\text{VI},z}^{\min}(Bi) = \begin{cases} 0.241 - 0.241e^{-Bi} & (0.001 \leq Bi < 0.1) \\ 0.427 - 0.165e^{-Bi} - 0.260e^{-\frac{Bi}{5}} & (0.1 \leq Bi \leq 10) \\ 0.467 - 0.124e^{-\frac{Bi}{10}} - 0.0338e^{-\frac{Bi}{50}} & (10 < Bi \leq 100) \end{cases}$ |
| Sphere | 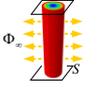 Case VII | |
| | $g_{\text{VII},\theta}^{\max}(Bi) = \begin{cases} 0.194 - 0.194e^{-Bi} & (0.001 \leq Bi < 0.1) \\ 0.5 - 0.0578e^{-Bi} - 0.43e^{-\frac{Bi}{5}} & (0.1 \leq Bi \leq 10) \\ 0.805 - 0.248e^{-\frac{Bi}{10}} - 0.318e^{-\frac{Bi}{50}} & (10 < Bi \leq 100) \end{cases}$ | $-g_{\text{VII},r}^{\min}(Bi) = -g_{\text{VII},\theta}^{\min}(Bi)$<br>$= \begin{cases} 0.193 - 0.193e^{-Bi} & (0.001 \leq Bi < 0.1) \\ 0.350 - 0.132e^{-Bi} - 0.217e^{-\frac{Bi}{5}} & (0.1 \leq Bi \leq 10) \\ 0.382 - 0.103e^{-\frac{Bi}{10}} - 0.0268e^{-\frac{Bi}{50}} & (10 < Bi \leq 100) \end{cases}$ |

It is found that the normalized DISes are all monotonically increasing function of the Biot number and approach 1 for the infinite $Bi$. The maximum stresses for the cases with free expansion are less than those with rigid in-plane constraint.

One interesting thing is that the maximum DISes of mode I and mode II for $Bi > 5$ are almost the same, while there is obvious difference between mode III and mode IV, although each pair has the same mechanical boundary conditions. The reason can be understood as follows. The larger $Bi = SH/D$ means the relative faster interface diffusion or slower diffusion in the bulk. In mode I and mode II, when the diffusion-induced stress reaches its maximum at a critical time, which is usually very soon after the initiation of the diffusion, the detectable diffusion front has not arrived at the bottom surface. Therefore the bottom boundary condition cannot significantly affect the maximum DIS. But in mode III and mode IV, the maximum diffusion-induced stress appears at infinite time, at that time the diffusion front has reached the bottom boundary, so the different bottom diffusion boundary will lead to different results.

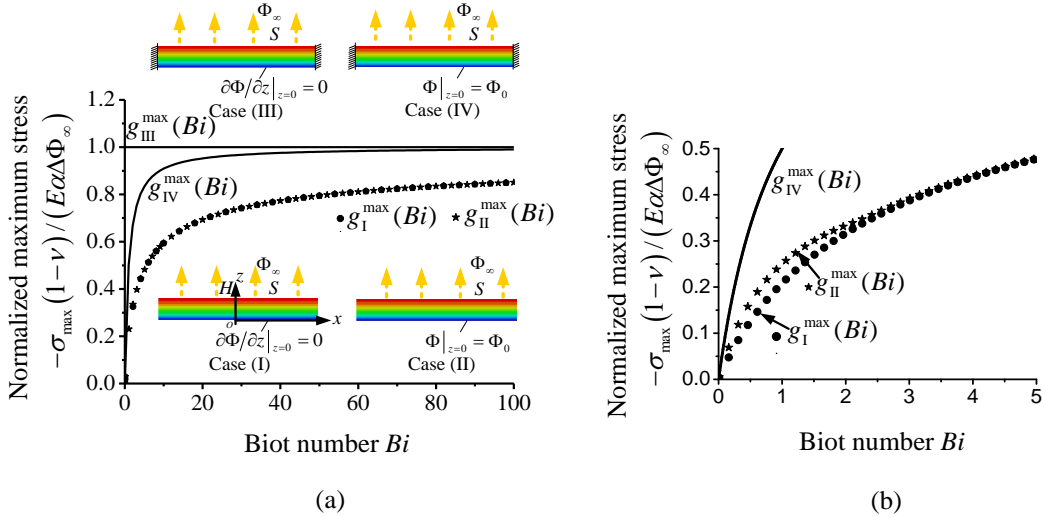

**Fig.7.** (a) Variation of normalized maximum thermal stresses as a function of the Biot number for four cases: Diffusion in a plate with free in-plane expansion and insulated boundary (Mode I); free in-plane expansion and isothermal boundary (Mode II); rigid in-plane constraint and insulated boundary (Mode III); rigid in-plane constraint and isothermal boundary (Mode IV). (b) a zoom-in view for small Biot number, to indicate that $g_\mathrm{I}^{\max}(Bi)$ and $g_\mathrm{II}^{\max}(Bi)$ are slightly different. The red color represents tensile stress and blue color means compressive stress.

## 4. The maximum diffusion induced stress of different geometers

Cylindrical or spherical geometries undergoing diffusion are often observed in real applications, such as the cylinder anodes in lithium battery subjected to the diffusion of lithium-ion. In this section, we study the effect of geometry factors on the diffusion-induced stress.

4.1 Diffusion in a cylinder
*Mode V Contraction Diffusion in a cylinder with fixed axial constraint*

Since the diffusion is developing along the radial direction, the diffusion governing equation in cylindrical coordinate system $(r, \theta, z)$ is

$$\frac{\partial \Phi}{\partial t} = D\left(\frac{\partial^2 \Phi}{\partial r^2} + \frac{1}{r}\frac{\partial \Phi}{\partial r}\right) \tag{31}$$

The corresponding boundary conditions are

$$D\frac{\partial \Phi}{\partial r}\bigg|_{r=R} = -S(\Phi|_{r=R} - \Phi_\infty) \tag{32}$$

$$\frac{\partial \Phi}{\partial r}\bigg|_{r=0} = 0 \tag{33}$$

where $R$ is the radius of the cylinder. The initial condition is

$$\Phi|_{t=0} = \Phi_0 \tag{34}$$

Using the dimensionless parameters $\hat{\Phi} = \frac{\Phi - \Phi_0}{\Phi_\infty - \Phi_0}$, $\hat{t} = \frac{Dt}{R^2}$, $\hat{r} = \frac{r}{R}$ and $Bi = \frac{SR}{D}$,

we obtain the solution to the equation in infinite series form as

$$\hat{\Phi}(\hat{r}, \hat{t}, Bi) = -\sum_{n=1}^{\infty} \frac{2\omega_n J_1(\omega_n) J_0(\omega_n \hat{r})}{(Bi^2 + \omega_n^2) J_0^2(\omega_n)} e^{-\omega_n^2 \hat{t}} + 1 \tag{35}$$

where $\omega_n$ is the n-th positive solution of

$$-\omega_n J_1(\omega_n) + Bi J_0(\omega_n) = 0 \tag{36}$$

$J_0$ and $J_1$ are the Bessel functions of the first kind.

The constitutive relationship in the cylindrical coordinate system are as the same as Eq. (14), and the subscript 1,2,3 correspond to $r$, $\theta$, $z$ respectively. The equilibrium condition is

$$\frac{d\sigma_r}{dr} + \frac{\sigma_r - \sigma_\theta}{r} = 0 \tag{37}$$

and the kinematic relations can be expressed as

$$\varepsilon_r = \frac{du_r}{dr}, \varepsilon_\theta = \frac{u_r}{r}, \varepsilon_z = 0 \tag{38}$$

where $u_r$ is the displacement in the radial direction. Combining Eq. (37), Eq. (38), and the constitutive relations, we can get

$$\frac{d^2 u_r}{dr^2} + \frac{1}{r}\frac{du_r}{dr} - \frac{1}{r^2}u_r = \frac{1+v}{1-v}\frac{d\varepsilon_\Phi}{dr} \tag{39}$$

Noting the following conditions

$$u_r|_{r=0} = 0 \tag{40}$$

$$\sigma_r|_{r=R} = 0 \tag{41}$$

the radial, hoop and axial stresses can be solved as

$$\sigma_r(r) = -\frac{E\alpha\Delta\Phi_\infty}{1-v}\hat{g}_{V,r}(\hat{r}, \hat{t}, Bi) \tag{42}$$

$$\sigma_\theta(r) = -\frac{E\alpha\Delta\Phi_\infty}{1-v}\hat{g}_{V,\theta}(\hat{r}, \hat{t}, Bi) \tag{43}$$

$$\sigma_z(r) = -\frac{E\alpha\Delta\Phi_\infty}{1-v}\hat{g}_{V,z}(\hat{r}, \hat{t}, Bi, v) \tag{44}$$

where

$$\hat{g}_{V,r}\left(\hat{r},\hat{t},Bi\right)=-\int_{0}^{1}\hat{\Phi}\left(\hat{r},\hat{t},Bi\right)\hat{r}d\hat{r}+\frac{1}{\hat{r}^{2}}\int_{0}^{\hat{r}}\hat{\Phi}\left(\hat{r},\hat{t},Bi\right)\hat{r}d\hat{r} \tag{45}$$

$$\hat{g}_{V,\theta}\left(\hat{r},\hat{t},Bi\right)=-\int_{0}^{1}\hat{\Phi}\left(\hat{r},\hat{t},Bi\right)\hat{r}d\hat{r}-\frac{1}{\hat{r}^{2}}\int_{0}^{\hat{r}}\hat{\Phi}\left(\hat{r},\hat{t},Bi\right)\hat{r}d\hat{r}+\hat{\Phi}\left(\hat{r},\hat{t},Bi\right) \tag{46}$$

$$\hat{g}_{V,z}\left(\hat{r},\hat{t},Bi,v\right)=-2v\int_{0}^{1}\hat{\Phi}\left(\hat{r},\hat{t},Bi\right)\hat{r}d\hat{r}+\hat{\Phi}\left(\hat{r},\hat{t},Bi\right) \tag{47}$$

are dimensionless function related to the normalized radial, hoop and axial stresses, respectively. It is found that $\hat{g}_{V,z}\left(\hat{r},\hat{t},Bi,v\right)$ has one more variable, i.e. the Poisson ratio $v$. Similar as before, the maximum DISes can then be written as

$$\sigma_{r}^{\max}=-\frac{E\alpha\Delta\Phi_{\infty}}{1-v}g_{V,r}^{\max}(Bi) \tag{48}$$

$$\sigma_{\theta}^{\max}=-\frac{E\alpha\Delta\Phi_{\infty}}{1-v}g_{V,\theta}^{\max}(Bi) \tag{49}$$

$$\sigma_{z}^{\max}=-\frac{E\alpha\Delta\Phi_{\infty}}{1-v}g_{V,z}^{\max}(Bi,v) \tag{50}$$

where $g_{V,r}^{\max}(Bi)$, $g_{V,\theta}^{\max}(Bi)$ and $g_{V,z}^{\max}(Bi,v)$ are obtained by maximizing $\hat{g}_{V,r}\left(\hat{r},\hat{t},Bi\right)$, $\hat{g}_{V,\theta}\left(\hat{r},\hat{t},Bi\right)$ and $\hat{g}_{V,z}\left(\hat{r},\hat{t},Bi,v\right)$ numerically, and the positive components are shown in Fig 8(a). Considering that only tensile stress causes fracture, $g_{V,r}^{\max}(Bi)\leq0$ (i.e. $\sigma_{z}^{\max}\leq0$) in the contraction diffusion process is therefore not shown.

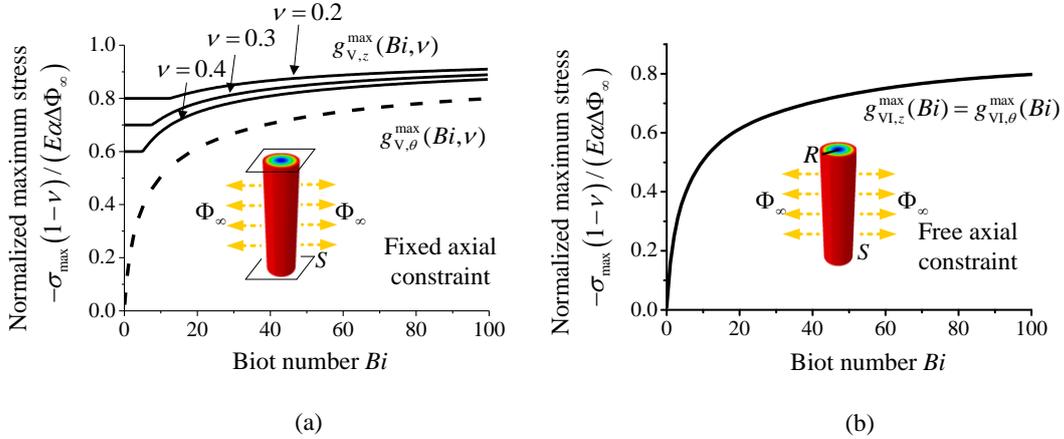

(a)          (b)

**Fig.8.** Variation of the normalized maximum diffusion induced stresses as functions of the Biot number (a) for mode V: diffusion in a cylinder with rigid axial constraint, (b) for mode VI: diffusion in a cylinder with free in-plane expansion. The red color represents tensile stress and blue color means compressive stress.

Among all the normalized maximum DIS components, $g_{V,z}^{\max}(Bi,v)$ is the largest. Meanwhile, we can find $g_{V,z}^{\max}(Bi,v)$ is reducing with the increasing of $v$. It is noted that there is a turning point on the curve of $g_{V,z}^{\max}(Bi,v)$ versus $Bi$, which can be understood as follows. If the Biot number $Bi$ is small, the maximum DIS appears after sufficient long time when the diffusion variable is homogeneous and equal to the value of the environment. If $Bi$ is large, the maximum DIS appears at the surface ($\hat{r}=1$) and time $\hat{t} = \hat{t}_{max}$ as we previously discussed in Section 2.2. The variations of DIS with time for different $Bi$ are illustrated by Fig 9. Hence, there exists a critical $Bi = SR/D$ (or a critical radius) between these two situations.

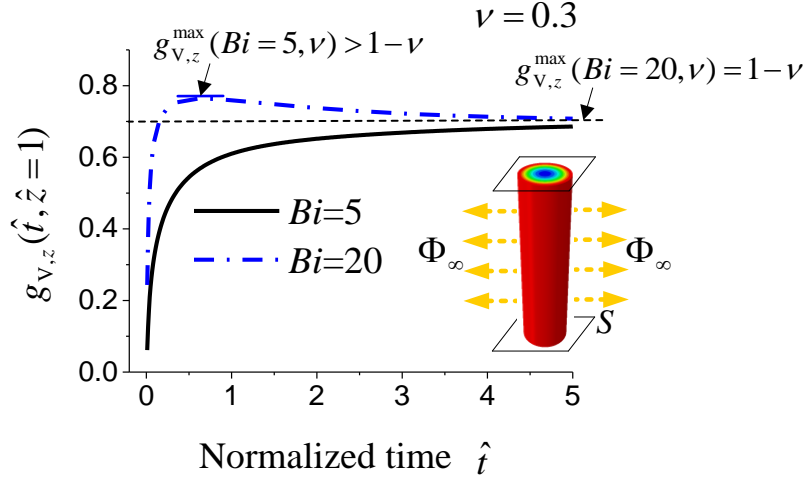

**Fig.9.** Variation of the normalized diffusion induced stress $g_{V,z}(\hat{t},\hat{z})$ (mode V, z direction) as a function of the normalized time for three different Biot numbers at the surface of the cylinder ($\hat{z}=1$). The red color represents tensile stress and blue color means compressive stress.

*Mode VI Diffusion in a cylinder with free axial expansion*

Different from the previous study on an axially constrained cylinder, the cylinder with free ends implies

$$\int_0^R \sigma_z(r,t) r dr = 0 \tag{51}$$

Replacing the condition $\varepsilon_z = 0$ by Eq. (51) and using the same equilibrium equation and kinetic relations as in mode V, we can get

$$\sigma_z = -\frac{E\alpha\Delta\Phi_\infty}{1-v} \hat{g}_{VI,z}(\hat{r},\hat{t},Bi) \tag{52}$$

where

$$\hat{g}_{VI,z}(\hat{r},\hat{t},Bi) = -2\int_0^1 \hat{\Phi}(\hat{r},\hat{t},Bi)\hat{r}d\hat{r} + \hat{\Phi}(\hat{r},\hat{t},Bi) \tag{53}$$

It is easy to know $\sigma_r, \sigma_\theta$ are the same as Eq. (42) and Eq. (43), and therefore

$$\sigma_r(r) = -\frac{E\alpha\Delta\Phi_\infty}{1-v} \hat{g}_{VI,r}(\hat{r},\hat{t},Bi) \tag{54}$$

$$\sigma_\theta(r) = -\frac{E\alpha\Delta\Phi_\infty}{1-v} \hat{g}_{VI,\theta}(\hat{r},\hat{t},Bi) \tag{55}$$

where

$$\hat{g}_{VI,r} = \hat{g}_{V,r}(\hat{r},\hat{t},Bi) \tag{56}$$

$$\hat{g}_{VI,\theta} = \hat{g}_{V,\theta}(\hat{r},\hat{t},Bi) \tag{57}$$

and

$$\sigma_r^{\max} = -\frac{E\alpha\Delta\Phi_\infty}{1-v} g_{VI,r}^{\max}(Bi) \tag{58}$$

$$\sigma_\theta^{\max} = -\frac{E\alpha\Delta\Phi_\infty}{1-v} g_{VI,\theta}^{\max}(Bi) \tag{59}$$

$$\sigma_z^{\max} = -\frac{E\alpha\Delta\Phi_\infty}{1-v} g_{VI,z}^{\max}(Bi) \tag{60}$$

are obtained by maximizing the DISes respectively.

The positive normalized DISes components for mode VI are all presented in Fig.8.(b), we can find that all the normalized DISes components are monotonically increasing function of the Biot number, and the maximum DIS along axial direction is less than that of the fixed axial condition and independent of Poisson ratio.

4.2. Diffusion in a sphere (mode VII)

Sometimes, diffusion occurs in particles. In this case, the diffusion is carrying on along the radial direction of a sphere. The diffusion equation in spherical coordinate system $(r,\theta,\varphi)$ can be presented as

$$\frac{\partial \Phi}{\partial t} = D\left(\frac{\partial^2 \Phi}{\partial r^2} + \frac{2}{r}\frac{\partial \Phi}{\partial r}\right) \tag{61}$$

The boundary conditions are

$$D\frac{\partial \Phi}{\partial r}\Big|_{r=R} = -S(\Phi|_{r=R} - \Phi_\infty) \tag{62}$$

$$\frac{\partial \Phi}{\partial r}\Big|_{r=0} = 0 \tag{63}$$

The initial condition is

$$\Phi|_{t=0} = \Phi_0 \tag{64}$$

Using the dimensionless parameters $\hat{\Phi} = \frac{\Phi - \Phi_0}{\Phi_\infty - \Phi_0}$, $\hat{t} = \frac{Dt}{R^2}$, $\hat{r} = \frac{r}{R}$ and $Bi = \frac{SR}{D}$, we can solve that

$$\hat{\Phi}(\hat{r},\hat{t},Bi) = -\sum_{n=1}^{\infty} \frac{4\sin(\omega_n \hat{r})(\sin\omega_n - \omega_n \cos\omega_n)}{2\hat{r}\omega_n^2 - \hat{r}\omega_n \sin(2\omega_n)} e^{-\omega_n^2 \hat{t}} + 1 \tag{65}$$

$\omega_n$ is the n-th positive solution of

$$-\sin\omega_n + \omega_n \cos\omega_n + Bi\sin\omega_n = 0 \tag{66}$$

Due to the symmetry of sphere, we can get $\sigma_\theta = \sigma_\varphi$ and $\varepsilon_\theta = \varepsilon_\varphi$. By corresponding 1,2,3 to $r$, $\theta$ and $\varphi$, the constitutive equations Eq. (14) become

$$\sigma_r = \frac{E}{(1+v)(1-2v)}\left[(1-2v)\varepsilon_r + v(\varepsilon_r + 2\varepsilon_\theta)\right] - \frac{E}{1-2v}\varepsilon_\Phi$$

$$\sigma_\theta = \frac{E}{(1+v)(1-2v)}\left[(1-2v)\varepsilon_\theta + v(\varepsilon_r + 2\varepsilon_\theta)\right] - \frac{E}{1-2v}\varepsilon_\Phi \tag{67}$$

The equilibrium condition in spherical coordinate system is

$$\frac{d\sigma_r}{dr} + 2\frac{\sigma_r - \sigma_\theta}{r} = 0 \tag{68}$$

and the kinematic relations can be expressed by

$$\varepsilon_r = \frac{du_r}{dr}, \varepsilon_\theta = \frac{u_r}{r} \tag{69}$$

where $u_r$ is the displacement in the radial direction. From Eqs. (67)-(69) and the boundary conditions

$$u_r \big|_{r=0} = 0 \tag{70}$$

$$\sigma_r \big|_{r=R} = 0 \tag{71}$$

we can relate the DISes to the diffusion variable as

$$\sigma_r(r) = -\frac{E\alpha\Delta\Phi_\infty}{1-v} \hat{g}_{VII,r}(\hat{r},\hat{t},Bi) \tag{72}$$

$$\sigma_\theta(r) = -\frac{E\alpha\Delta\Phi_\infty}{1-v} \hat{g}_{VII,\theta}(\hat{r},\hat{t},Bi) \tag{73}$$

where

$$\hat{g}_{VII,r}(\hat{r},\hat{t},Bi) = -2\int_0^1 \hat{\Phi}(\hat{r},\hat{t},Bi)\hat{r}^2 d\hat{r} + \frac{2}{\hat{r}^3}\int_0^{\hat{r}} \hat{\Phi}(\hat{r},\hat{t},Bi)\hat{r}^2 d\hat{r} \tag{74}$$

$$\hat{g}_{VII,\varphi}(\hat{r},\hat{t},Bi) = \hat{g}_{VII,\theta}(\hat{r},\hat{t},Bi) = -2\int_0^1 \hat{\Phi}(\hat{r},\hat{t},Bi)\hat{r}^2 d\hat{r} - \frac{1}{\hat{r}^3}\int_0^{\hat{r}} \hat{\Phi}(\hat{r},\hat{t},Bi)\hat{r}^2 d\hat{r} + \hat{\Phi}(\hat{r},\hat{t},Bi) \tag{75}$$

The maximum normalized DISes can then be obtained by maximizing Eq. (74) and (75), and are expressed by

$$\sigma_r^{max} = \frac{-E\alpha\Delta\Phi_\infty}{(1-v)} g_{VII,r}^{max}(Bi)$$

$$\sigma_\theta^{max} = \frac{-E\alpha\Delta\Phi_\infty}{(1-v)} g_{VII,\theta}^{max}(Bi) \tag{76}$$

$$\sigma_\varphi^{max} = \frac{-E\alpha\Delta\Phi_\infty}{(1-v)} g_{VII,\varphi}^{max}(Bi)$$

We find that $g_{VII,r}^{max}(Bi)$ is negative which does not cause fracture, and $g_{VII,\theta}^{max}(Bi) = g_{VII,\varphi}^{max}(Bi)$ is plotted in Fig 10. It is found from Fig 10 that decreasing the Biot number $Bi = SR/D$ through reducing the radius of particles $R$, is an effective way to reduce DISes in materials such as in lithium battery system.

The maximum DISes of the rectangular plane, the cylinder and the sphere are compared in Fig 11, and their average curvatures are 0, 0.5/$R$ and 1/$R$, respectively. It is found that the curvature can influence the maximum DISes. With the average curvature increases, the in-plane constraint for expansion or contraction becomes weaker, so the DISes decrease.

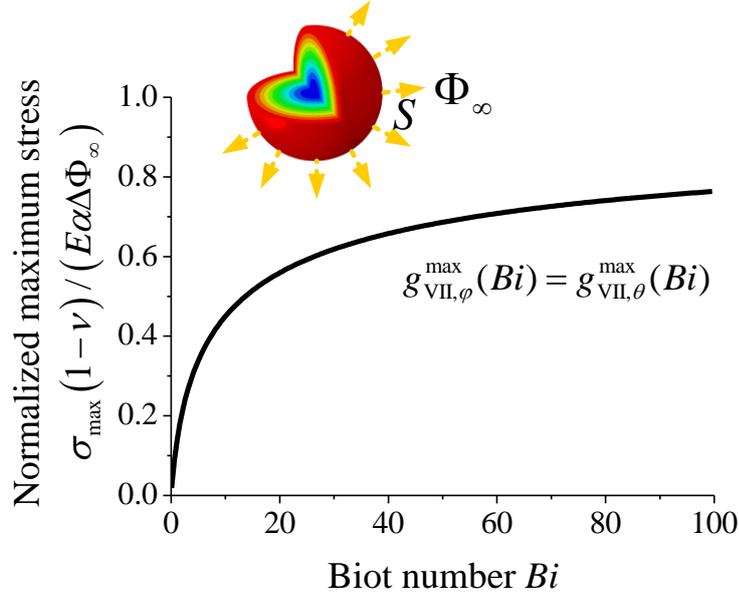

**Fig.10.** Variation of normalized maximum diffusion induced stresses as functions of the Biot number for case VII: diffusion in a sphere (Mode VII). The red color represents tensile stress and blue color means compressive stress.

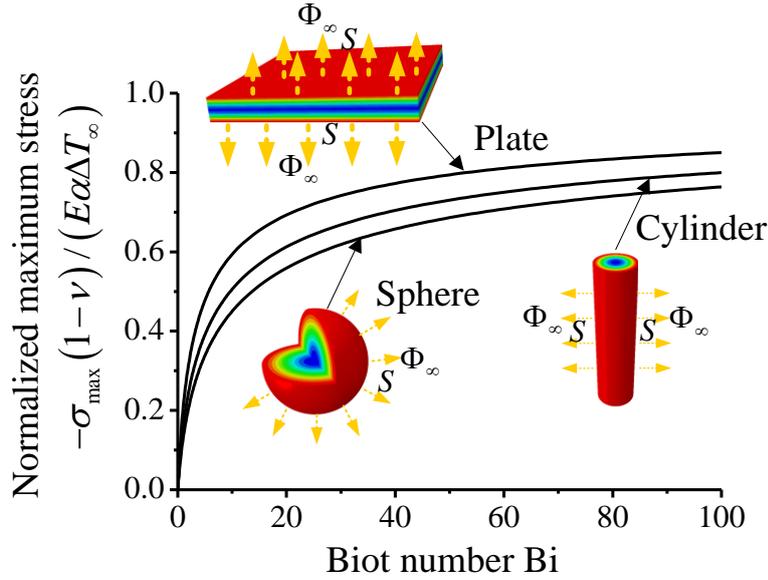

**Fig.11.** The compares of the normalized maximum diffusion induced stresses varying with the Biot number beteween plate, cylinder and sphere in free constraint. The red color represents tensile stress and blue color means compressive stress.

## 5. The maximum diffusion induced stress for the stress-dependent diffusion

To model the DIS more precisely, the coupling effect of stress on the diffusion process should not be ignored in some cases. The case studied here is the delithiation in lithium batteries. Considering the coupling effect, the species flux **J** is a function of the stress and the concentration gradient of the diffusing species. Therefore, the governing equation of the diffusion process becomes nonlinear and more complex. For example, in an infinite plate (the inset of Fig. 2), the equation is

$$\frac{\partial \Phi(z,t)}{\partial t} = \frac{\partial \left[ D\left(1 + \frac{2\Omega E\alpha}{3(1-v)R_g T}\Delta\Phi(z,t)\right) \frac{\partial \Phi(z,t)}{\partial z} \right]}{\partial z} \tag{77}$$

where $R_g$ is the gas constant, $T$ is the absolute temperature, $X$ is the mole fraction, $\Omega$ is the partial molar volume of diffusion species. The derivation to obtain Eq. (77) is presented in Appendix B. Eq. (77) can be normalized as

$$\frac{\partial \hat{\Phi}(\hat{z},\hat{t})}{\partial \hat{t}} = \frac{\partial \left[ \left(1 + \hat{\beta}\hat{\Phi}\right) \frac{\partial \hat{\Phi}(\hat{z},\hat{t})}{\partial \hat{z}} \right]}{\partial \hat{z}} \tag{78}$$

by using the normalized variables or parameters $\hat{\Phi} = \frac{\Delta\Phi}{\Delta\Phi_\infty}$, $\hat{t} = \frac{Dt}{H^2}$, $\hat{z} = \frac{z}{H}$, and $\hat{\beta} = \frac{2\Omega E\alpha \Phi_\infty}{3(1-v)R_g T}$. From Eq. (78), we note that the coupling effect can be reflected by the normalized parameter $\hat{\beta}$, which is named as the coupling coefficient in this paper. The boundary conditions of mode I with the coupling effect are

$$D\left(1 + \frac{2\Omega E\alpha}{3(1-v)R_g T}\Delta\Phi\right)\frac{\partial \Phi}{\partial z}\bigg|_{z=H} = -S(\Phi|_{z=H} - \Phi_\infty) \tag{79}$$

$$\frac{\partial \Phi}{\partial z}\bigg|_{z=0} = 0 \tag{80}$$

where $D = MR_g T$ is the diffusion coefficient and $M$ is the mobility of lithium ions. The initial condition is

$$\Phi|_{t=0} = 0 \tag{81}$$

Eqs. (79)-(81) can be rewritten in the normalized form as

$$\left(1 + \hat{\beta}\hat{\Phi}\right)\frac{\partial \hat{\Phi}}{\partial \hat{z}}\bigg|_{\hat{z}=1} = -Bi(\hat{\Phi}|_{\hat{z}=1} - 1) \tag{82}$$

$$\frac{\partial \hat{\Phi}}{\partial \hat{z}}\bigg|_{\hat{z}=0} = 0 \tag{83}$$

$$\hat{\Phi}|_{\hat{t}=0} = 0 \tag{84}$$

Solving Eq. (78) with the boundary/initial conditions Eqs. (82)-(84), we can find that the normalized diffusion variable can be expressed as

$$\hat{\Phi} = \hat{\Phi}\left(\hat{t}, \hat{z}, Bi, \hat{\beta}\right) \tag{85}$$

Similar to Eq. (18), we introduce a normalized DIS,

$$\hat{G}_I\left(\hat{z},\hat{t},Bi,\hat{\beta}\right) = \frac{\sigma_{xx}(z,t)}{-E\alpha\Phi_\infty/(1-v)} = \hat{\Phi}\left(\hat{z},\hat{t},Bi,\hat{\beta}\right) - \int_0^1 \hat{\Phi}\left(\hat{z},\hat{t},Bi,\hat{\beta}\right)d\hat{z} \tag{86}$$

In this coupling case, the maximum DIS is only related to $Bi$ and $\hat{\beta}$, i.e.,

$$\sigma^{max} = -\frac{E\alpha\Delta\Phi_\infty}{1-v}G_I^{max}(Bi,\hat{\beta}) \tag{87}$$

where $G_I^{max}(Bi,\hat{\beta})$ is the maximum of $\hat{G}_I\left(\hat{z},\hat{t},Bi,\hat{\beta}\right)$ with respect to $\hat{z}$ and $\hat{t}$.

To obtain $G_\mathrm{I}^{\max}(Bi,\hat{\beta})$, the first step is to obtain the diffusion variable $\hat{\Phi}(\hat{z},\hat{t})$. $\hat{\Phi}(\hat{z},\hat{t})$ can be accurately calculated by numerical methods such as the finite difference algorithm. When the coupling coefficient $\hat{\beta}$ is small ($\hat{\beta}<1$), the diffusion variable $\hat{\Phi}$ can also be solved as a power series of $\hat{\beta}$ using the perturbation method (See Appendix C). Once $\hat{\Phi}$ is obtained, $G_\mathrm{I}^{\max}(Bi,\hat{\beta})$ is calculated by maximizing Eq. (86).

For other boundary conditions and geometries (cylinder and sphere), the maximum DISes can be obtained following the same procedure as above. Here we only list the corresponding governing equations and boundary/initial conditions for the cylinder and the sphere case respectively.

Cylinder (in cylindrical coordinate system):

$$\frac{\partial \hat{\Phi}}{\partial \hat{t}} = \frac{1}{\hat{r}}\frac{\partial}{\partial \hat{r}}\left[\hat{r}\left(1+\hat{\beta}\hat{\Phi}\right)\frac{\partial \hat{\Phi}}{\partial \hat{r}}\right] \tag{88}$$

$$\left(1+\hat{\beta}\hat{\Phi}\right)\frac{\partial \hat{\Phi}}{\partial \hat{r}}\bigg|_{\hat{r}=1} = -Bi(\hat{\Phi}|_{\hat{r}=1}-1) \tag{89}$$

$$\frac{\partial \hat{\Phi}}{\partial \hat{r}}\bigg|_{\hat{r}=0} = 0 \tag{90}$$

$$\hat{\Phi}\big|_{\hat{t}=0} = 0 \tag{91}$$

Sphere (in spherical coordinate system):

$$\frac{\partial \hat{\Phi}}{\partial \hat{t}} = \frac{1}{\hat{r}^2}\frac{\partial}{\partial \hat{r}}\left[\hat{r}^2\left(1+\hat{\beta}\hat{\Phi}\right)\frac{\partial \hat{\Phi}}{\partial \hat{r}}\right] \tag{92}$$

$$\left(1+\hat{\beta}\hat{\Phi}\right)\frac{\partial \hat{\Phi}}{\partial \hat{r}}\bigg|_{\hat{r}=1} = -Bi(\hat{\Phi}|_{\hat{r}=1}-1) \tag{93}$$

$$\frac{\partial \hat{\Phi}}{\partial \hat{r}}\bigg|_{\hat{r}=0} = 0 \tag{94}$$

$$\hat{\Phi}\big|_{\hat{t}=0} = 0 \tag{95}$$

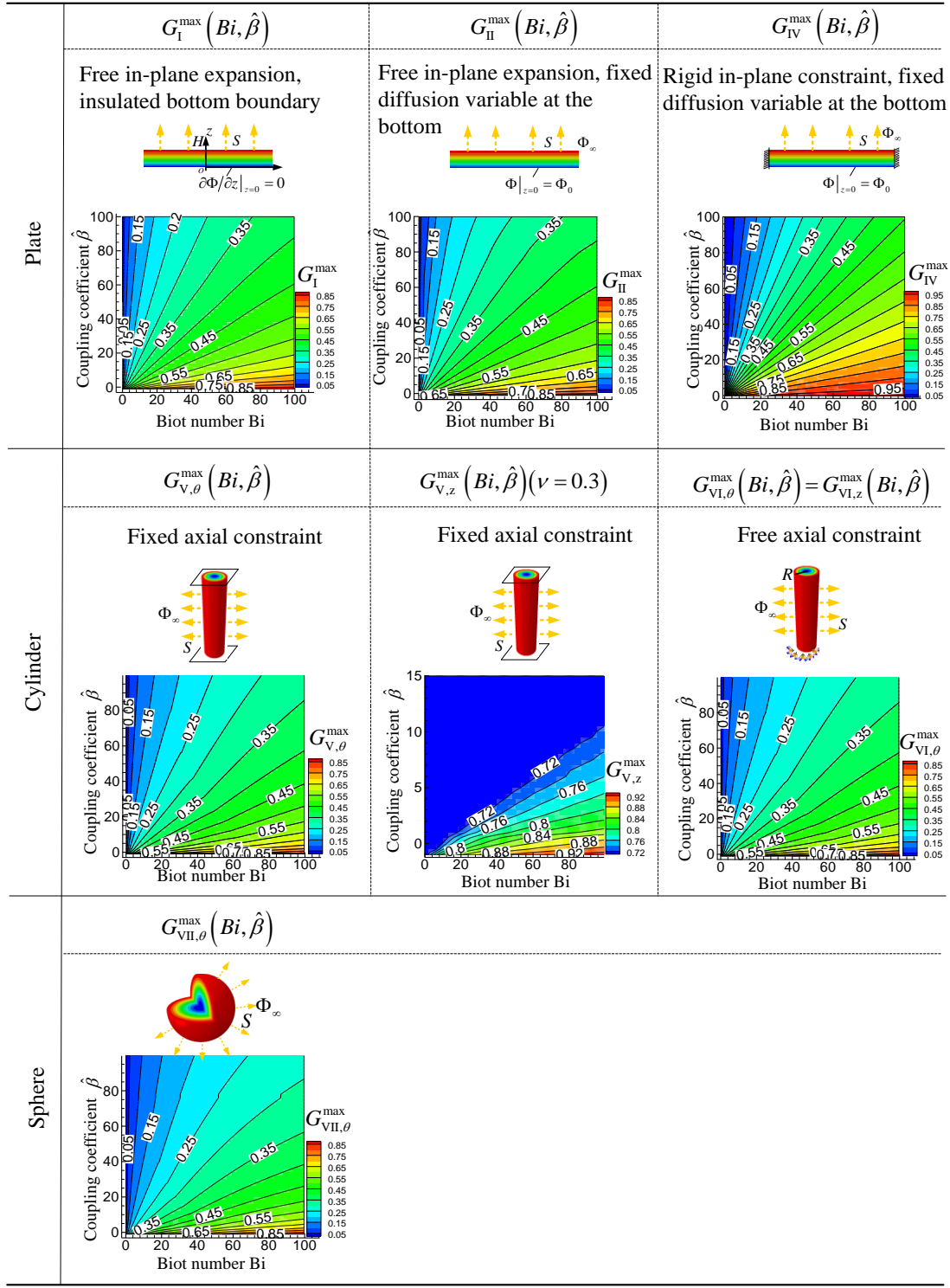

**Fig.12.** Variation of normalized maximum diffusion induced stresses as a function of the Biot number Bi and the coupling coefficient $\hat{\beta}$ for various cases. The case of fixed in-plane constraint and fixed diffusion variable at the bottom (mode III) is not shown because $G_{\mathrm{III}}^{\max}=1$ regardless of the value of $Bi$ or $\hat{\beta}$.

All the positive normalized maximum DISes of various cases are shown in Fig. 12 except for $G_{\text{III}}^{\max}$ which always equals 1. $G_{\text{IV}}^{\max}$ can be obtained analytically as

$$G_{\text{IV}}^{\max}(Bi,\hat{\beta}) = \frac{Bi}{1+Bi+\hat{\beta}} \qquad (96)$$

while $G_{\text{II}}^{\max}(Bi,\hat{\beta})$, $G_{\text{V},\theta}^{\max}(Bi,\hat{\beta})$, $G_{\text{V},z}^{\max}(Bi,\hat{\beta})$, $G_{\text{VI},\theta}^{\max}(Bi,\hat{\beta})$, $G_{\text{VI},z}^{\max}(Bi,\hat{\beta})$, $G_{\text{VII},\theta}^{\max}(Bi,\hat{\beta})$ and $G_{\text{VII},\varphi}^{\max}(Bi,\hat{\beta})$ are calculated numerically. We can find that the normalized maximum DISes are all reducing with the increasing of $\hat{\beta}$ except $G_{\text{III}}^{\max}(Bi,\hat{\beta})$. The reason is because that the tensile stress enhances the diffusion in the material and the concentration of the diffusing species becomes more homogeneous during the process, resulting in reduced stresses. $G_{\text{III}}^{\max}(Bi,\hat{\beta})$ is an exception, because it is independent of the diffusion process and equals 1. The results also indicate that if we neglect the fully stress-diffusion coupling effect, the relative error will exceed 10% when $\hat{\beta}$ is larger than 1 (see Fig C1 (b) in the appendix). As we have discussed, the coupling coefficient $\hat{\beta}$ in the diffusion process of the lithium battery can be far larger than 1 in some cases and therefore the fully stress-diffusion coupling effect plays an important role and must be considered in the theoretical model.

From Fig. 12, we observe that when $Bi$ and $\hat{\beta}$ are large, the contour lines of the maximum DISes are close to rays through the origin, indicating that $Bi$ and $\hat{\beta}$ can be further combined into one parameter approximately. This can be understood if we rewrite the governing equation Eq. (77) and the boundary condition Eq. (79) as

$$\frac{\partial \hat{\Phi}(\hat{z},\hat{t})}{\partial \hat{t}} = \frac{\partial^2 \hat{\Phi}(\hat{z},\hat{t})}{\partial \hat{z}^2} \qquad (97)$$

$$\left.\frac{\partial \hat{\Phi}}{\partial \hat{z}}\right|_{\hat{z}=1} = -\tilde{Bi}(\hat{\Phi}|_{\hat{z}=1} - 1) \qquad (98)$$

where $\hat{t} = \frac{\tilde{D}t}{H^2}$, $\hat{z} = \frac{z}{H}$, $\tilde{Bi} = \frac{SH}{\tilde{D}} = \frac{Bi}{1+\hat{\beta}\hat{\Phi}}$ and $\tilde{D} = D(1+\hat{\beta}\hat{\Phi})$ is the equivalent diffusion coefficient. Eq. (97) and Eq. (98) have the same form with their counterpart for stress-independent diffusion cases, i.e. Eq (10) and Eq. (11), and therefore the maximum DISes are dependent only on the equivalent Biot number $\tilde{Bi}$. The maximum DISes can be fitted as functions of $\tilde{Bi} = \frac{Bi}{1+\eta\hat{\beta}} = \frac{3SH(1-v)R_g T}{D+2D\eta\Omega E\alpha\Phi_\infty}$ as presented by Table 2, where $\eta$ is fitted as 0.4. When $\tilde{Bi} \geq 10$, the relative error of the approximate fitting functions are less than 10% for all the cases studied above.

**Table 2.** The approximate fitting functions of the maximum diffusion induced stresses in stress-dependent diffusion

| | | |
|---|---|---|
| Plate | Case I<br>Free in-plane expansion, insulated bottom boundary 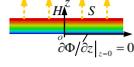 | |
| | $G_{\mathrm{I}}^{\max}(Bi,\hat{\beta}) = \begin{cases} \left(0.617 - 0.156 e^{-\frac{Bi}{1+0.4\hat{\beta}}} - 0.351 e^{-\frac{1}{51+0.4\hat{\beta}}\frac{Bi}{}}\right) & \left(0.1 \leq \frac{Bi}{1+0.4\hat{\beta}} \leq 10\right) \\ \left(0.890 - 0.221 e^{-\frac{1}{101+0.4\hat{\beta}}\frac{Bi}{}} - 0.279 e^{-\frac{1}{501+0.4\hat{\beta}}\frac{Bi}{}}\right) & \left(10 \leq \frac{Bi}{1+0.4\hat{\beta}} \leq 100\right) \end{cases}$ | |
| | Case II<br>Free in-plane expansion, fixed diffusion variable at the bottom 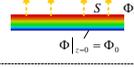 | |
| | $G_{\mathrm{II}}^{\max}(Bi,\hat{\beta}) = \begin{cases} \left(0.610 - 0.169 e^{-\frac{Bi}{1+0.4\hat{\beta}}} - 0.340 e^{-\frac{1}{51+0.4\hat{\beta}}\frac{Bi}{}}\right) & \left(0.1 \leq \frac{Bi}{1+0.4\hat{\beta}} \leq 10\right) \\ \left(0.908 - 0.222 e^{-\frac{1}{101+0.4\hat{\beta}}\frac{Bi}{}} - 0.300 e^{-\frac{1}{501+0.4\hat{\beta}}\frac{Bi}{}}\right) & \left(10 \leq \frac{Bi}{1+0.4\hat{\beta}} \leq 100\right) \end{cases}$ | |
| Cylinder | Case V<br>Fixed axial constraint 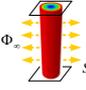 | |
| | $G_{\mathrm{V},\theta}^{\max}(Bi,\hat{\beta}) = \begin{cases} \left(0.538 - 0.139 e^{-\frac{Bi}{1+0.4\hat{\beta}}} - 0.322 e^{-\frac{1}{51+0.4\hat{\beta}}\frac{Bi}{}}\right) & \left(0.1 \leq \frac{Bi}{1+0.4\hat{\beta}} \leq 10\right) \\ \left(0.887 - 0.141 e^{-\frac{1}{101+0.4\hat{\beta}}\frac{Bi}{}} - 0.395 e^{-\frac{1}{501+0.4\hat{\beta}}\frac{Bi}{}}\right) & \left(10 \leq \frac{Bi}{1+0.4\hat{\beta}} \leq 100\right) \end{cases}$ | |
| | $G_{\mathrm{V},z}^{\max}(Bi,\hat{\beta},\nu=0.30) = \max\left(1-\nu, F(Bi,\hat{\beta})\right) \quad (\nu=0.30)$<br><br>$F(Bi,\hat{\beta}) = \begin{cases} \left(0.766 - 0.010 e^{-\frac{Bi}{1+0.4\hat{\beta}}} - 0.340 e^{-\frac{1}{51+0.4\hat{\beta}}\frac{Bi}{}}\right) & \left(0.1 \leq \frac{Bi}{1+0.4\hat{\beta}} \leq 10\right) \\ \left(0.947 - 0.028 e^{-\frac{1}{101+0.4\hat{\beta}}\frac{Bi}{}} - 0.317 e^{-\frac{1}{501+0.4\hat{\beta}}\frac{Bi}{}}\right) & \left(10 \leq \frac{Bi}{1+0.4\hat{\beta}} \leq 100\right) \end{cases}$ | |
| | Case VI<br>Free axial constraint 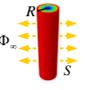 | |
| | $G_{\mathrm{VI},\theta}^{\max}(Bi) = G_{\mathrm{VI},z}^{\max}(Bi) = G_{\mathrm{V},\theta}^{\max}(Bi)$ | |
| Sphere | Case VII 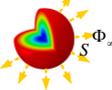 | |
| | $G_{\mathrm{VII},\theta}^{\max}(Bi,\hat{\beta}) = \begin{cases} \left(0.490 - 0.111 e^{-\frac{Bi}{1+0.4\hat{\beta}}} - 0.311 e^{-\frac{1}{51+0.4\hat{\beta}}\frac{Bi}{}}\right) & \left(0.1 \leq \frac{Bi}{1+0.4\hat{\beta}} \leq 10\right) \\ \left(0.870 - 0.109 e^{-\frac{1}{101+0.4\hat{\beta}}\frac{Bi}{}} - 0.442 e^{-\frac{1}{501+0.4\hat{\beta}}\frac{Bi}{}}\right) & \left(10 \leq \frac{Bi}{1+0.4\hat{\beta}} \leq 100\right) \end{cases}$ | |

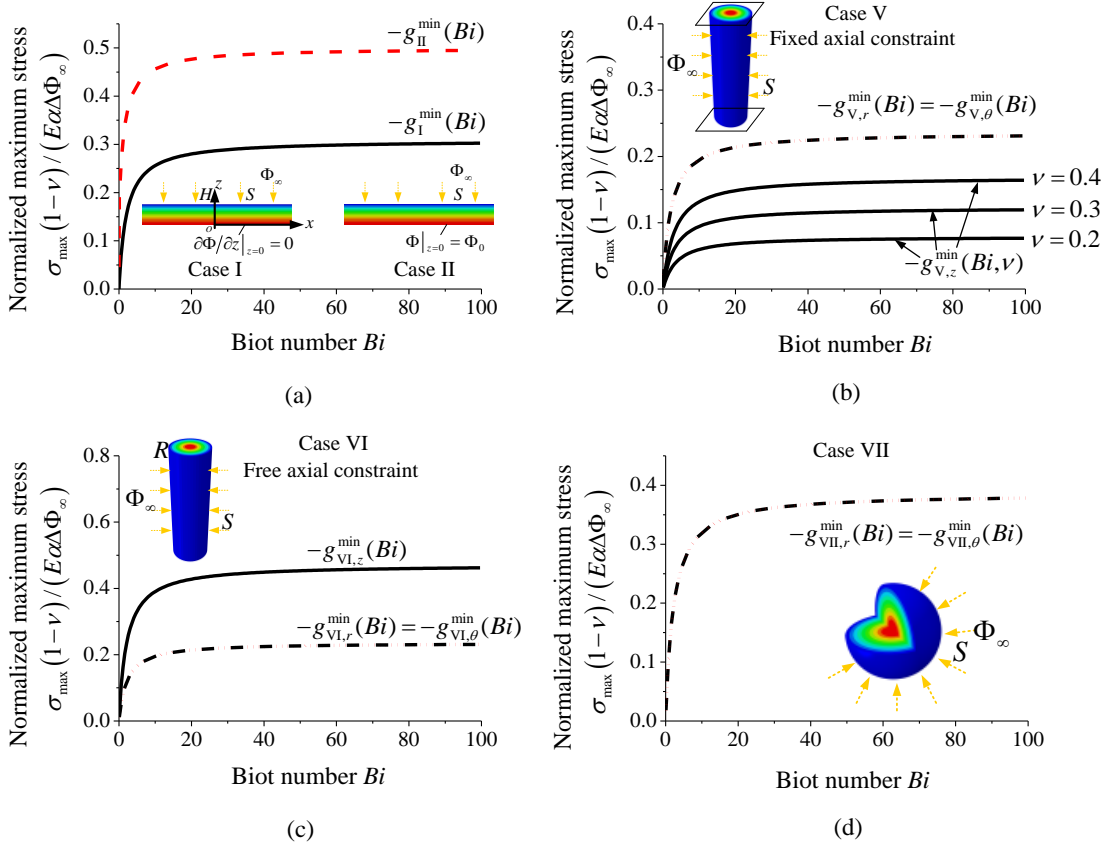

**Fig 13**. Variation of normalized maximum thermal stresses as a function of the Biot number for (a) a plate with free in-plane expansion, (b) a cylinder with rigid longitudinal constraint, (c) a cylinder with free longitudinal expansion and (d) a sphere with free expansion.

## 6. The maximum expansion-diffusion induced stress

In previous sections, the maximum contraction diffusion induced stress and fracture are investigated. In this section, we consider the case where the material expands due to heating or inward diffusing of species, i.e. $\alpha\Delta\Phi_\infty > 0$. As we have assumed that fracture is induced by tensile stress, one difference between contraction diffusion and expansion diffusion is that the maximum tensile stress emerges at the surfaces in the former case and at the inner region of the materials in the latter case.

According to Eq. (18) the maximum expansion DIS can be obtained by minimizing $\hat{g}(\hat{z},\hat{t},Bi)$. Therefore, the normalized maximum expansion DIS is

$$\frac{\sigma^{\max}(1-\nu)}{E\alpha\Delta\Phi_\infty} = -g_I^{\min}(Bi) = -\min(\hat{g}(\hat{z},\hat{t},Bi)) \qquad (99)$$

Similarly, the normalized maximum expansion DISes for other cases can be obtained as $-g_{II}^{\min}(Bi)$, $-g_{V,r}^{\min}(Bi)$, $-g_{V,\theta}^{\min}(Bi)$, $-g_{V,z}^{\min}(Bi,\nu)$, $-g_{VI,r}^{\min}(Bi)$, $-g_{VI,\theta}^{\min}(Bi)$, $-g_{VI,z}^{\min}(Bi)$, $-g_{VII,r}^{\min}(Bi)$, $-g_{VII,\theta}^{\min}(Bi)$ by minimizing their expressions in Eqs. (A34)(A51)(A52)(A53)(A54)(A55)(A56)(A73)(A74), respectively. The results of plate, cylinder and sphere are drawn in Fig 13. Under the same magnitude of $\Delta\Phi_\infty$, the maximum DISes in expansion diffusion are smaller than those of contraction diffusion in most cases, implying that materials are relatively safer. This is because that the largest tensile stress occurs at the inner region of the material, rather than the surface as in the contraction diffusion case. The heat or diffusing species need some time to reach the inner region, and during this period the material could homogenize its temperature difference within it. Therefore the deformation is more homogenous and the stress is smaller.

**7. Conclusions**

In summary, we develop quantitative models for various diffusion processes which can successfully predict the maximum diffusion induced stress, the critical concentration difference at which the fracture occurs, the crack density and the hierarchical crack patterns observed in experiments. Simple and accurate formulae are obtained for engineers and materialists to quickly calculate the maximum diffusion induced stress for various typical geometries and working conditions. They are in closed analytical form except a single-parameter term for stress-independent diffusion or a two-parameter term for stress-dependent diffusion which is determined by numerical methods and presented by curves or contour plots. More conclusions can be drawn as follows

(1) The crack density induced by diffusion is almost a constant, independent of the temperature difference (or the concentration difference) and the thickness of specimen. When the temperature difference further increases beyond another threshold, hierarchy crack patterns appear because part of the cracks further propagate. The phenomena may guide the design of the lithium ion battery and the thermal protection system. For example, if the size of platelet is smaller than the diffusion induced crack-spacing, no fracture will happen anymore.

(2) Another potential application of our analytical fracture model is that the fracture toughness of ceramics can be determined by alternatively measuring the thermal shock induced crack-spacing and the strength.

(3) The in-plane constraint for expansion or contraction plays an important role on the maximum DISes. It is influenced by the boundary conditions, the curvature and the Biot number (related to the thickness of specimens). The corresponding quantitative relations have been obtained, which implies that a specimen with smaller thickness or radius can sustain more dramatic diffusion processes safely.

(4) When the dependence of the diffusion process on the stress cannot be ignored, one more normalized coupling parameter $\hat{\beta}$ is necessary in predicting the maximum DISes. This parameter can combine with the Biot number $Bi$ to form an equivalent Biot number $\tilde{Bi}$ appearing in the approximate formulae on the maximum DISes. The resulting relative error due this approximation is less than 10% with a wide range on the $Bi$-$\hat{\beta}$ plane.

(5) The maximum contraction-DIS emerges on the surface at very beginning with a dramatic gradient of the diffusion variable, while the maximum expansion-DIS emerges at the inner of materials after a relatively longer period with a mild gradient of the diffusion variable. Therefore, the former has lager value and is more dangerous.

**Acknowledgement**

The authors acknowledge the support from National Natural Science Foundation of China (Grant Nos. 11372158, 11090334, and 51232004), National Basic Research

Program of China (973 Program Grant No. 2010CB832701), and Tsinghua University Initiative Scientific Research Program (No. 2011Z02173).Program of China (973 Program Grant No. 2010CB832701), and Tsinghua University Initiative Scientific Research Program (No. 2011Z02173).

**Reference**

Bahr, H.-A., Weiss, H.-J., Bahr, U., Hofmann, M., Fischer, G., Lampenscherf, S., Balke, H., 2010. Scaling behavior of thermal shock crack patterns and tunneling cracks driven by cooling or drying. Journal of the Mechanics and Physics of Solids 58, 1411-1421.

Bahr, H.A., Fischer, G., Weiss, H.J., 1986. Thermal-Shock Crack Patterns Explained by Single and Multiple Crack-Propagation. J Mater Sci 21, 2716-2720.

Bhandakkar, T.K., Gao, H.J., 2010. Cohesive modeling of crack nucleation under diffusion induced stresses in a thin strip: Implications on the critical size for flaw tolerant battery electrodes. Int J Solids Struct 47, 1424-1434.

Bohn, S., Douady, S., Couder, Y., 2005. Four sided domains in hierarchical space dividing patterns. Phys Rev Lett 94.

Bower, A.F., Guduru, P.R., 2012. A simple finite element model of diffusion, finite deformation, plasticity and fracture in lithium ion insertion electrode materials. Model Simul Mater Sc 20.

Cheng, Y.T., Verbrugge, M.W., 2010a. Application of Hasselman's Crack Propagation Model to Insertion Electrodes. Electrochem Solid St 13, A128-A131.

Cheng, Y.T., Verbrugge, M.W., 2010b. Diffusion-Induced Stress, Interfacial Charge Transfer, and Criteria for Avoiding Crack Initiation of Electrode Particles. J Electrochem Soc 157, A508-A516.

Christensen, J., Newman, J., 2006. Stress generation and fracture in lithium insertion materials. J Solid State Electr 10, 293-319.

Collin, M., Rowcliffe, D., 2002. The morphology of thermal cracks in brittle materials. J Eur Ceram Soc 22, 435-445.

Deshpande, R., Cheng, Y.T., Verbrugge, M.W., Timmons, A., 2011. Diffusion Induced Stresses and Strain Energy in a Phase-Transforming Spherical Electrode Particle. J Electrochem Soc 158, A718-A724.

Erdogan, F., Ozturk, M., 1995. Periodic Cracking of Functionally Graded Coatings. Int J Eng Sci 33, 2179-2195.

Fahrenholtz, W.G., Hilmas, G.E., Talmy, I.G., Zaykoski, J.A., 2007. Refractory diborides of zirconium and hafnium. J Am Ceram Soc 90, 1347-1364.

Hasselman, D.P.H, 1969. Unified Theory of Thermal Shock Fracture Initiation and Crack Propagation in Brittle Ceramics. J Am Ceram Soc 52, 600-604.

Huggins, R.A., Nix, W.D., 2000. Decrepitation Model For Capacity Loss During Cycling of Alloys in Rechargeable Electrochemical Systems. Ionics 6, 57-63.

Hutchinson, J.W., Xia, Z.C., 2000. Crack patterns in thin films. Journal of the Mechanics and Physics of Solids 48, 1107-1131.

Lei, H.J., Liu, B., Wang, C.A., Fang, D.N., 2012. Study on biomimetic staggered composite for better thermal shock resistance. Mech Mater 49, 30-41.

Levine, S.R., Opila, E.J., Halbig, M.C., Kiser, J.D., Singh, M., Salem, J.A., 2002. Evaluation of ultra-high temperature ceramics for aeropropulsion use. J Eur Ceram Soc 22, 2757-2767.

Li, J.C., Dozier, A.K., Li, Y.C., Yang, F.Q., Cheng, Y.T., 2011. Crack Pattern Formation in Thin Film Lithium-Ion Battery Electrodes. J Electrochem Soc 158, A689-A694.

Lim, C., Yan, B., Yin, L.L., Zhu, L.K., 2012. Simulation of diffusion-induced stress using reconstructed electrodes particle structures generated by micro/nano-CT. Electrochim Acta 75, 279-287.

Lim, M.R., Cho, W.I., Kim, K.B., 2001. Preparation and characterization of gold-codeposited LiMn2O4 electrodes. J Power Sources 92, 168-176.
Liu, H., Wu, Y.S., Lambropoulos, J.C., 2009. Thermal shock and post-quench strength of lapped borosilicate optical glass. J Non-Cryst Solids 355, 2370-2374.
Lu, T.J., Fleck, N.A., 1998. The thermal shock resistance of solids. Acta Mater 46, 4755-4768.
Ma, B.X., Han, W.B., 2010. Thermal shock resistance of ZrC matrix ceramics. Int J Refract Met H 28, 187-190.
Manson, S.S., 1954. Behaviour of materials under conditions of thermal stress. Nat. Advis. Commun.Aeromaut. Rep. No.1, 170.
Monteverde, F., Guicciardi, S., Melandri, C., Fabbriche, D.D., 2010. Densification, Microstructure Evolution and Mechanical Properties of Ultrafine Sic Particle-Dispersed Zrb2 Matrix Composites. Nato Sec Sci B Phys, 261-272.
Park, J., Lu, W., Sastry, A.M., 2011. Numerical Simulation of Stress Evolution in Lithium Manganese Dioxide Particles due to Coupled Phase Transition and Intercalation. J Electrochem Soc 158, A201-A206.
Purkayastha, R.T., McMeeking, R.M., 2012. An integrated 2-D model of a lithium ion battery: the effect of material parameters and morphology on storage particle stress. Comput Mech 50, 209-227.
Qi, Y., Harris, S.J., 2010. In Situ Observation of Strains during Lithiation of a Graphite Electrode. J Electrochem Soc 157, A741-A747.
Ryu, I., Choi, J.W., Cui, Y., Nix, W.D., 2011. Size-dependent fracture of Si nanowire battery anodes. Journal of the Mechanics and Physics of Solids 59, 1717-1730.
Shi, D.H., Xiao, X.R., Huang, X.S., Kia, H., 2011. Modeling stresses in the separator of a pouch lithium-ion cell. J Power Sources 196, 8129-8139.
Song, F., Meng, S.H., Xu, X.H., Shao, Y.F., 2010. Enhanced Thermal Shock Resistance of Ceramics through Biomimetically Inspired Nanofins. Phys Rev Lett 104.
Swain, M.V., 1990. R-Curve Behavior and Thermal-Shock Resistance of Ceramics. J Am Ceram Soc 73, 621-628.
Thackeray, M.M., Shao-Horn, Y., Kahaian, A.J., Kepler, K.D., Vaughey, J.T., Hackney, S.A., 1998. Structural fatigue in spinel electrodes in high voltage (4V) Li/LixMn2O4 cells. Electrochem Solid St 1, 7-9.
Tucker, M.C., Kroeck, L., Reimer, J.A., Cairns, E.J., 2002. The influence of covalence on capacity retention in metal-substituted spinels - Li-7 NMR, SQUID, and electrochemical studies. J Electrochem Soc 149, A1409-A1413.
Vanimisetti, S.K., Ramakrishnan, N., 2012. Effect of the electrode particle shape in Li-ion battery on the mechanical degradation during charge-discharge cycling. P I Mech Eng C-J Mec 226, 2192-2213.
Verbrugge, M.W., Cheng, Y.T., 2009. Stress and Strain-Energy Distributions within Diffusion-Controlled Insertion-Electrode Particles Subjected to Periodic Potential Excitations. J Electrochem Soc 156, A927-A937.
Woodford, W.H., Carter, W.C., Chiang, Y.M., 2012. Design criteria for electrochemical shock resistant battery electrodes. Energ Environ Sci 5, 8014-8024.
Woodford, W.H., Chiang, Y.M., Carter, W.C., 2010. "Electrochemical Shock" of Intercalation Electrodes: A Fracture Mechanics Analysis. J Electrochem Soc 157, A1052-A1059.
Xiao, X., Liu, P., Verbrugge, M.W., Haftbaradaran, H., Gao, H., 2011. Improved cycling stability of silicon thin film electrodes through patterning for high energy density lithium batteries. J Power Sources 196, 1409-1416.
Zhang, X.C., Shyy, W., Sastry, A.M., 2007. Numerical simulation of intercalation-

induced stress in Li-ion battery electrode particles. J Electrochem Soc 154, A910-A916.
Zhao, K.J., Tritsaris, G.A., Pharr, M., Wang, W.L., Okeke, O., Suo, Z.G., Vlassak, J.J., Kaxiras, E., 2012. Reactive Flow in Silicon Electrodes Assisted by the Insertion of Lithium. Nano Lett 12, 4397-4403.

Appendix A: Analytic solution of the diffusion variables and stresses
A.1. Mode I: plate with free in-plane expansion and fixed diffusion variable at the bottom

The governing equations and initial/boundary conditions are Eqs. (10)-(13). As the first step, we let

$$\hat{\Psi} = \hat{\Phi} - 1 \tag{A1}$$

Therefore $\hat{\Psi}$ satisfies

$$\frac{\partial \hat{\Psi}}{\partial \hat{t}} = \frac{\partial^2 \hat{\Psi}}{\partial \hat{z}^2} \tag{A2}$$

$$\left.\frac{\partial \hat{\Psi}}{\partial \hat{z}}\right|_{\hat{z}=1} = -Bi\,\hat{\Psi}\big|_{\hat{z}=1} \tag{A3}$$

$$\left.\frac{\partial \hat{\Psi}}{\partial \hat{z}}\right|_{\hat{z}=0} = 0 \tag{A4}$$

$$\hat{\Psi}\big|_{\hat{t}=0} = -1 \tag{A5}$$

Use the separation of variable techinique, let

$$\hat{\Psi} = Z(\hat{z})T(\hat{t}) \tag{A6}$$

Substitute Eq. (A6) into Eq. (A2), we obtain

$$Z'' + \lambda Z = 0 \tag{A7}$$
$$T' + \lambda T = 0 \tag{A8}$$

For $\lambda < 0$, the general solution to Eq. (A7) is

$$Z = Ae^{-\sqrt{\lambda}\hat{z}} + Be^{\sqrt{\lambda}\hat{z}} \tag{A9}$$

The boundary conditions Eq. (A3) and Eq. (A4) lead to

$$A\left(Bi - \sqrt{\lambda}\right)e^{-\sqrt{\lambda}} + B\left(Bi + \sqrt{\lambda}\right)e^{\sqrt{\lambda}} = 0 \tag{A10}$$

$$-A + B = 0 \tag{A11}$$

It is obvious from the above two equations that $A = B = 0$, giving rise to the trivial solution $Z = 0$. For $\lambda = 0$, we have

$$Z = A\hat{z} + B \tag{A12}$$

and the boundary conditions again require $A = B = 0$. When $\lambda = \omega^2 > 0$, the general solution to Eq. (A7) reads

$$Z = A\cos(\omega\hat{z}) + B\sin(\omega\hat{z}) \tag{A13}$$

Consider the boundary conditions Eq. (A3) and Eq. (A4), we have

$$Z'(0) = 0, Z'(1) + BiZ(1) = 0 \tag{A14}$$

Combine Eq. (A13) and Eq. (A14), we have

$$B = 0 \tag{A15}$$

$$\omega \tan \omega = Bi \tag{A16}$$

The eigenvalues $\omega_n$ are then solved from Eq. (A16)(Only the positive roots are taken because $-\omega \tan(-\omega) = \omega \tan \omega$ and $B\sin(-\omega\hat{z}) = -B\sin(\omega\hat{z})$).

Therefore the solution can be written in the following series form

$$\hat{\Psi} = \sum_{n=1}^{\infty} A_n e^{-\omega_n^2 \hat{t}} \cos(\omega_n \hat{z}) \tag{A17}$$

Using the initial condition Eq. (A5), we have

$$\sum_{n=1}^{\infty} A_n \cos(\omega_n \hat{z}) = -1 \tag{A18}$$

By expanding the right side of Eq. (A18), we obtain the coefficient $A_n$ as

$$A_n = -\frac{1}{L_n} \int_0^1 \cos(\omega_n \hat{z}) \mathrm{d}\hat{z} = -\frac{1}{L_n \omega_n} \sin \omega_n \tag{A19}$$

where

$$L_n = \int_0^1 \cos^2(\omega_n \hat{z}) \mathrm{d}\hat{z} = \frac{1}{2} + \frac{1}{4\omega_n} \sin(2\omega_n) \tag{A20}$$

Therefore the diffusion variable and the DIS are

$$\hat{\Phi} = -\sum_{n=1}^{\infty} \frac{4 \sin \omega_n \cos(\omega_n \hat{z})}{2\omega_n + \sin(2\omega_n)} e^{-\omega_n^2 \hat{t}} + 1 \tag{A21}$$

$$\hat{g}_I(\hat{z}, \hat{t}, Bi) = \sum_{n=1}^{\infty} \frac{4 \sin^2 \omega_n - 4\omega_n \sin \omega_n \cos(\omega_n \hat{z})}{2\omega_n^2 + \omega_n \sin(2\omega_n)} e^{-\omega_n^2 \hat{t}} \tag{A22}$$

A.2. Mode II: plate with free in-plane expansion and fixed diffusion variable at the bottom

Compared to the case in Section A.1, we only need to replace the boundary condition Eq. (12) by the following condition

$$\hat{\Phi}\big|_{\hat{z}=0} = 0 \tag{A23}$$

Let

$$\hat{\Psi} = \hat{\Phi} - \frac{Bi}{1+Bi} \hat{z} \tag{A24}$$

Then $\hat{\Psi}$ satisfies

$$\frac{\partial \hat{\Psi}}{\partial \hat{t}} = \frac{\partial^2 \hat{\Psi}}{\partial \hat{z}^2} \tag{A25}$$

$$\frac{\partial \hat{\Psi}}{\partial \hat{z}}\bigg|_{\hat{z}=1} = -Bi\, \hat{\Psi}\big|_{\hat{z}=1} \tag{A26}$$

$$\hat{\Psi}\big|_{\hat{z}=0} = 0 \tag{A27}$$

$$\hat{\Psi}\big|_{\hat{t}=0} = -\frac{Bi}{1+Bi} \hat{z} \tag{A28}$$

Using the separation of variable technique similar to Section A.1, we obtain the eigenfunctions as $e^{-\omega_n^2 \hat{t}} \sin(\omega_n \hat{z})$, where the eigenvalue $\omega_n$ is the n-th positive root of

$$\tan \omega = -\frac{1}{Bi}\omega \tag{A29}$$

The solution is written as

$$\hat{\Psi} = \sum_{n=1}^{\infty} A_n e^{-\omega_n^2 \hat{t}} \sin(\omega_n \hat{z}) \tag{A30}$$

Applying the initial condition Eq. (A28), the coefficient $A_n$ is solved as

$$A_n = -\frac{Bi}{1+Bi}\frac{1}{L_n}\int_0^1 \hat{z}\sin(\omega_n \hat{z})d\hat{z} = -\frac{Bi}{1+Bi}\frac{1}{L_n}\left(-\frac{1}{\omega_n}\cos\omega_n + \frac{1}{\lambda_n}\sin\omega_n\right) \tag{A31}$$

where

$$L_n = \int_0^1 \sin^2(\omega_n \hat{z})d\hat{z} = \frac{1}{2} - \frac{1}{4\omega_n}\sin 2\omega_n \tag{A32}$$

Therefore the expression for the diffusion variable is

$$\hat{\Phi} = -\sum_{n=1}^{\infty}\frac{Bi}{1+Bi}\frac{\sin\omega_n - \omega_n\cos\omega_n}{2\omega_n^2 - \omega_n\sin 2\omega_n}\sin(\omega_n \hat{z})e^{-\omega_n^2 \hat{t}} + \frac{Bi}{1+Bi}z \tag{A33}$$

The expression for the DIS is

$$\hat{g}_{\mathrm{II}}(\hat{z},\hat{t},Bi) = -\sum_{n=1}^{\infty}\frac{Bi}{1+Bi}\frac{\sin\omega_n - \omega_n\cos\omega_n}{2\omega_n^3 - \omega_n^2\sin 2\omega_n}(\cos\omega_n - 1)e^{-\omega_n^2\hat{t}} - \frac{Bi}{2(1+Bi)}$$
$$-\sum_{n=1}^{\infty}\frac{Bi}{1+Bi}\frac{\sin\omega_n - \omega_n\cos\omega_n}{2\omega_n^2 - \omega_n\sin 2\omega_n}\sin(\omega_n \hat{z})e^{-\omega_n^2\hat{t}} + \frac{Bi}{1+Bi}z \tag{A34}$$

A.3. Mode V: Diffusion in a cylinder with fixed axial constraint

The governing Equation and the initial/boundary conditions are Eqs.(31)-(34), which are normalized as

$$\frac{\partial \hat{\Phi}}{\partial \hat{t}} = \frac{\partial^2 \hat{\Phi}}{\partial \hat{r}^2} + \frac{1}{\hat{r}}\frac{\partial \hat{\Phi}}{\partial \hat{r}} \tag{A35}$$

$$\frac{\partial \hat{\Phi}}{\partial \hat{r}}\bigg|_{\hat{r}=1} = -Bi(\hat{\Phi}\big|_{\hat{r}=1} - 1) \tag{A36}$$

$$\frac{\partial \Phi}{\partial \hat{r}}\bigg|_{\hat{r}=0} = 0 \tag{A37}$$

$$\hat{\Phi}\big|_{\hat{t}=0} = 0 \tag{A38}$$

Let $\hat{\Psi} = \hat{\Phi} - 1$, which satisfies

$$\frac{\partial \hat{\Psi}}{\partial \hat{t}} = \frac{\partial^2 \hat{\Psi}}{\partial \hat{r}^2} + \frac{1}{\hat{r}}\frac{\partial \hat{\Psi}}{\partial \hat{r}} \tag{A39}$$

$$\frac{\partial \hat{\Psi}}{\partial \hat{r}}\bigg|_{\hat{r}=1} = -Bi\hat{\Psi}\big|_{\hat{r}=1} \tag{A40}$$

$$\frac{\partial \hat{\Psi}}{\partial \hat{r}}\bigg|_{\hat{r}=0} = 0 \tag{A41}$$

$$\hat{\Psi}\big|_{\hat{t}=0} = -1 \tag{A42}$$

Employing the separation of variable method and letting $\hat{\Psi} = R(\hat{r})T(\hat{t})$, we obtain

$$R''(\hat{r}) + \frac{1}{\hat{r}}R'(\hat{r}) + \omega^2 R(\hat{r}) = 0 \tag{A43}$$

$$T'(\hat{t}) + \omega^2 T(\hat{t}) = 0 \tag{A44}$$

The general solution to Eq. (A43) is

$$R_n = C_n J_0(\omega_n \hat{r}) \tag{A45}$$

where $J_0$ is the Bessel function with order zero, and the boundary condition at $\hat{r} = 1$ requests

$$-\omega_n J_1(\omega_n) + Bi J_0(\omega_n) = 0 \tag{A46}$$

where $J_1$ is the Bessel function with order one. The eigenvalues are therefore solved from Eq. (A46) (taken the positive roots only). The solution is written in the form

$$\hat{\Psi} = \sum_{n=1}^{\infty} A_n J_0(\omega_n \hat{r}) e^{-\omega_n^2 \hat{t}} \tag{A47}$$

Applying the initial condition Eq. (A42), we can solve that

$$A_n = -\frac{1}{N_n^2} \int_0^1 \hat{r} J_0(\omega_n \hat{r}) d\hat{r} = -\frac{1}{N_n^2 \omega_n^2} \omega_n J_1(\omega_n) \tag{A48}$$

where

$$N_n^2 = \int_0^1 \hat{r} J_0^2(\omega_n \hat{r}) d\hat{r} = \frac{1}{2\omega_n^2}\left(Bi^2 + \omega_n^2\right) J_0^2(\omega_n) \tag{A49}$$

Therefore the diffusion variable and the DISes are finally obtained as

$$\hat{\Phi}(\hat{r},\hat{t},Bi) = -\sum_{n=1}^{\infty} \frac{2\omega_n J_1(\omega_n) J_0(\omega_n \hat{r})}{(Bi^2 + \omega_n^2) J_0^2(\omega_n)} e^{-\omega_n^2 t} + 1 \tag{A50}$$

$$\hat{g}_{V,r}(\hat{r},\hat{t},Bi) = \sum_{n=1}^{\infty} \frac{2\hat{r} J_1^2(\omega_n) - 2 J_1(\omega_n) J_1(\hat{r}\omega_n)}{(Bi^2 + \omega_n^2) \hat{r} J_0^2(\omega_n)} e^{-\omega_n^2 t} \tag{A51}$$

$$\hat{g}_{V,\theta}(\hat{r},\hat{t},Bi) = \sum_{n=1}^{\infty} \frac{2\hat{r} J_1^2(\omega_n) + 2 J_1(\omega_n) J_1(\hat{r}\omega_n) - 2\omega_n J_1(\omega_n) J_0(\omega_n \hat{r})}{(Bi^2 + \omega_n^2) \hat{r} J_0^2(\omega_n)} e^{-\omega_n^2 t}$$

$$\tag{A52}$$

$$\hat{g}_{V,z}(\hat{r},\hat{t},Bi) = 1 - v + \sum_{n=1}^{\infty} \frac{4v J_1^2(\omega_n) - 2\omega_n J_1(\omega_n) J_0(\omega_n \hat{r})}{(Bi^2 + \omega_n^2) J_0^2(\omega_n)} e^{-\omega_n^2 t} \tag{A53}$$

A.4. Mode VI: Diffusion in a cylinder with free axial expansion

The diffusion variable is the same as Section A.3. The DISes are

$$\hat{g}_{VI,r}(\hat{r},\hat{t},Bi) = \sum_{n=1}^{\infty} \frac{2\hat{r} J_1^2(\omega_n) - 2 J_1(\omega_n) J_1(\hat{r}\omega_n)}{(Bi^2 + \omega_n^2) \hat{r} J_0^2(\omega_n)} e^{-\omega_n^2 t} \tag{A54}$$

$$\hat{g}_{VI,\theta}(\hat{r},\hat{t},Bi) = \sum_{n=1}^{\infty} \frac{2\hat{r} J_1^2(\omega_n) + 2 J_1(\omega_n) J_1(\hat{r}\omega_n) - 2\omega_n J_1(\omega_n) J_0(\omega_n \hat{r})}{(Bi^2 + \omega_n^2) \hat{r} J_0^2(\omega_n)} e^{-\omega_n^2 t}$$

$$\tag{A55}$$

$$\hat{g}_{VI,z} = -\sum_{n=1}^{\infty} \frac{4 J_1^2(\omega_n) - 2\omega_n J_1(\omega_n) J_0(\omega_n \hat{r})}{(Bi^2 + \omega_n^2) J_0^2(\omega_n)} e^{-\omega_n^2 t} \tag{A56}$$

A.5. Mode VII: Diffusion in a sphere

The governing Equation and the initial/boundary conditions are Eqs.(61)-(64), which

are normalized as

$$\frac{\partial \hat{\Phi}}{\partial \hat{t}} = \frac{\partial^2 \hat{\Phi}}{\partial \hat{r}^2} + \frac{2}{\hat{r}} \frac{\partial \hat{\Phi}}{\partial \hat{r}} \tag{A57}$$

$$\left.\frac{\partial \hat{\Phi}}{\partial \hat{r}}\right|_{\hat{r}=1} = -Bi(\hat{\Phi}|_{\hat{r}=1} - 1) \tag{A58}$$

$$\left.\frac{\partial \hat{\Phi}}{\partial \hat{r}}\right|_{\hat{r}=0} = 0 \tag{A59}$$

$$\hat{\Phi}\big|_{\hat{t}=0} = 0 \tag{A60}$$

Let $\hat{\Psi} = \hat{\Phi} - 1$, which satisfies

$$\frac{\partial \hat{\Psi}}{\partial \hat{t}} = \frac{\partial^2 \hat{\Psi}}{\partial \hat{r}^2} + \frac{2}{\hat{r}} \frac{\partial \hat{\Psi}}{\partial \hat{r}} \tag{A61}$$

$$\left.\frac{\partial \hat{\Psi}}{\partial \hat{r}}\right|_{\hat{r}=1} = -Bi\hat{\Psi}\big|_{\hat{r}=1} \tag{A62}$$

$$\left.\frac{\partial \hat{\Psi}}{\partial \hat{r}}\right|_{\hat{r}=0} = 0 \tag{A63}$$

$$\hat{\Psi}\big|_{\hat{t}=0} = -1 \tag{A64}$$

Employing the separation of variable method and letting $\hat{\Psi} = R(\hat{r})T(\hat{t})$, we obatin

$$R''(\hat{r}) + \frac{2}{\hat{r}} R'(\hat{r}) + \omega^2 R(\hat{r}) = 0 \tag{A65}$$

$$T'(\hat{t}) + \omega^2 T(\hat{t}) = 0 \tag{A66}$$

The general solution to Eq. (A65) is

$$R_n = C_n \frac{1}{\omega_n \hat{r}} \sin(\omega_n \hat{r}) \tag{A67}$$

The boundary condition at $\hat{r} = 1$ requests

$$-\sin \omega_n + \omega_n \cos \omega_n + Bi \sin \omega_n = 0 \tag{A68}$$

The eigenvalues are therefore solved from Eq. (A68) (taken the positive roots only). The solution is then written in the form

$$\hat{\Psi} = \sum_{n=1}^{\infty} A_n \frac{1}{\omega_n \hat{r}} \sin(\omega_n \hat{r}) e^{-\omega_n^2 \hat{t}} \tag{A69}$$

Applying the initial condition Eq. (A64), we can solve that

$$A_n = -\frac{1}{N_n^2} \int_0^1 \frac{\hat{r}}{\omega_n} \sin(\omega_n \hat{r}) d\hat{r} = -\frac{1}{N_n^2} \left( -\frac{1}{\omega_n^2} \cos \omega_n + \frac{1}{\omega_n^3} \sin \omega_n \right) \tag{A70}$$

where

$$N_n^2 = \int_0^1 \frac{1}{\omega_n^2} \sin^2(\omega_n \hat{r}) d\hat{r} = \frac{1}{\omega_n^2} \left[ \frac{1}{2} - \frac{1}{4\omega_n} \sin(2\omega_n) \right] \tag{A71}$$

Therefore the diffusion variable and the DISes are

$$\hat{\Phi}(\hat{r}, \hat{t}, Bi) = -\sum_{n=1}^{\infty} \frac{4\sin(\omega_n \hat{r})(\sin \omega_n - \omega_n \cos \omega_n)}{2\hat{r}\omega_n^2 - \hat{r}\omega_n \sin(2\omega_n)} e^{-\omega_n^2 \hat{t}} + 1 \tag{A72}$$

$$\hat{g}_{\text{VII},r}\left(\hat{r},\hat{t},Bi\right) = \sum_{n=1}^{\infty}\frac{8\left(\sin\omega_n - \omega_n\cos\omega_n\right)^2}{2\omega_n^4 - \omega_n^3\sin(2\omega_n)}e^{-\omega_n^2 t}$$
$$-\sum_{n=1}^{\infty}\frac{8\left(\sin\omega_n - \omega_n\cos\omega_n\right)\left[\sin(\omega_n\hat{r}) - \hat{r}\omega_n\cos(\omega_n\hat{r})\right]}{2\hat{r}^3\omega_n^4 - \hat{r}^3\omega_n^3\sin(2\omega_n)}e^{-\omega_n^2 t}$$
(A73)

$$\hat{g}_{\text{VII},\theta}\left(\hat{r},\hat{t},Bi\right) = \sum_{n=1}^{\infty}\frac{8\left(\sin\omega_n - \omega_n\cos\omega_n\right)^2}{2\omega_n^4 - \omega_n^3\sin(2\omega_n)}e^{-\omega_n^2 t} - \sum_{n=1}^{\infty}\frac{4\sin(\omega_n\hat{r})\left(\sin\omega_n - \omega_n\cos\omega_n\right)}{2\hat{r}\omega_n^2 - \hat{r}\omega_n\sin(2\omega_n)}e^{-\omega_n^2 t}$$
$$+\sum_{n=1}^{\infty}\frac{8\left(\sin\omega_n - \omega_n\cos\omega_n\right)\left[\sin(\omega_n\hat{r}) - \hat{r}\omega_n\cos\omega_n\right]}{2\hat{r}^3\omega_n^4 - \hat{r}^3\omega_n^3\sin(2\omega_n)}e^{-\omega_n^2 t}$$
(A74)

$$\hat{g}_{\text{VII},\varphi}\left(\hat{r},\hat{t},Bi\right) = \hat{g}_{\text{VII},\theta}\left(\hat{r},\hat{t},Bi\right) \tag{A75}$$

Appendix B. Derivation of Eq. (77)

The species flux can be expressed as
$$\mathbf{J} = -M\Phi\nabla\mu \tag{B76}$$
where $M$ is the mobility of lithium ions, and $\mu$ is the electrochemical potential. The electrochemical potential in an ideal solid solution is
$$\mu = \mu_0 + R_g T \ln X - \Omega\sigma_H \tag{B77}$$
where
$$\sigma_H = \frac{\sigma_{11} + \sigma_{22} + \sigma_{33}}{3} \tag{B78}$$
and $\mu_0$ is a constant. Substituting Eq. (B78) into Eq. (B77) by noting that
$$\nabla\left(R_g T \ln X\right) = R_g T \frac{1}{X}\nabla X = R_g T \frac{1}{\Phi/c}\nabla(\Phi/c) = R_g T \frac{1}{\Phi}\nabla\Phi \tag{B79}$$
we can obtain
$$\mathbf{J} = -D\left(\nabla\Phi - \frac{\Omega\Phi}{R_g T}\nabla\sigma_H\right) \tag{B80}$$
where $D = MR_g T$ is the diffusion coefficient. To determine $\sigma_H$, we need to know the stress distribution. Here we take Mode I as an example (free in-plane expansion), where by considering $\sigma_{xx} = \sigma_{yy}, \sigma_{zz} = 0$, and Eq. (18), we have
$$\sigma_H = \frac{\left(\sigma_{xx} + \sigma_{yy} + \sigma_{zz}\right)}{3} = \frac{2E\alpha}{3(1-v)}\left[\frac{1}{H}\int_0^H \Delta\Phi(z,t)dz - \Delta\Phi(z,t)\right] \tag{B81}$$
and
$$\frac{\partial\sigma_H}{\partial z} = -\frac{2E\alpha}{3(1-v)}\frac{\partial\Phi(z,t)}{\partial z} \tag{B82}$$
Substituting Eq. (B82) into Eq. (B80) yields
$$J = -D\left(1 + \frac{2\Omega E\alpha}{3(1-v)R_g T}\Delta\Phi\right)\frac{\partial\Phi(z,t)}{\partial z} \tag{B83}$$

Considering Eq (2), we obtain the governing equation of diffusion process Eq. (77).

Appendix C: The perturbation solution to obtain the diffusion variable when $\hat{\beta}$ is small

We expand the diffusion variable $\hat{\Phi}$ as a power series of the coupling coefficient $\hat{\beta}$, namely

$$\hat{\Phi}(\hat{z},\hat{t}) = \hat{\Phi}_0(\hat{z},\hat{t}) + \hat{\beta}\hat{\Phi}_1(\hat{z},\hat{t}) + \hat{\beta}^2\hat{\Phi}_2(\hat{z},\hat{t}) + \hat{\beta}^3\hat{\Phi}_3(\hat{z},\hat{t}) + \ldots \quad (C1)$$

Substitute Eq. (C1) into Eq. (78) and Eqs.(82)-(84), and by equaling the coefficients with the same order of $\hat{\beta}$, we obtain

$$\begin{cases} \dfrac{\partial \hat{\Phi}_0(\hat{z},\hat{t})}{\partial \hat{t}} = \dfrac{\partial^2 \hat{\Phi}_0(\hat{z},\hat{t})}{\partial \hat{z}^2} \\ \left.\dfrac{\partial \hat{\Phi}_0}{\partial \hat{z}}\right|_{\hat{z}=1} = -Bi(\hat{\Phi}_0|_{\hat{z}=1} - 1) \\ \left.\dfrac{\partial \hat{\Phi}_0}{\partial \hat{z}}\right|_{\hat{z}=0} = 0 \\ \hat{\Phi}_0|_{\hat{t}=0} = 0 \end{cases} \quad (C2)$$

$$\begin{cases} \dfrac{\partial \hat{\Phi}_1(\hat{z},\hat{t})}{\partial \hat{t}} = \dfrac{\partial^2 \hat{\Phi}_1(\hat{z},\hat{t})}{\partial \hat{z}^2} + \dfrac{\partial}{\partial \hat{z}}\left[\hat{\Phi}_0(\hat{z},\hat{t})\dfrac{\partial \hat{\Phi}_0(\hat{z},\hat{t})}{\partial \hat{z}}\right] \\ \left.\dfrac{\partial \hat{\Phi}_1}{\partial \hat{z}}\right|_{\hat{z}=1} = -Bi\hat{\Phi}_1|_{\hat{z}=1} - \left(\hat{\Phi}_0 \dfrac{\partial \hat{\Phi}_0}{\partial \hat{z}}\right)\bigg|_{\hat{z}=1} \\ \left.\dfrac{\partial \hat{\Phi}_1}{\partial \hat{z}}\right|_{\hat{z}=0} = 0 \\ \hat{\Phi}_1|_{\hat{t}=0} = 0 \end{cases} \quad (C3)$$

$$\begin{cases} \dfrac{\partial \hat{\Phi}_2(\hat{z},\hat{t})}{\partial \hat{t}} = \dfrac{\partial^2 \hat{\Phi}_2(\hat{z},\hat{t})}{\partial \hat{z}^2} + \dfrac{\partial}{\partial \hat{z}}\left[\hat{\Phi}_1(\hat{z},\hat{t})\dfrac{\partial \hat{\Phi}_0(\hat{z},\hat{t})}{\partial \hat{z}} + \hat{\Phi}_0(\hat{z},\hat{t})\dfrac{\partial \hat{\Phi}_1(\hat{z},\hat{t})}{\partial \hat{z}}\right] \\ \left.\dfrac{\partial \hat{\Phi}_2}{\partial \hat{z}}\right|_{\hat{z}=1} = -Bi\hat{\Phi}_2|_{\hat{z}=1} - \left(\hat{\Phi}_1 \dfrac{\partial \hat{\Phi}_0}{\partial \hat{z}} + \hat{\Phi}_0 \dfrac{\partial \hat{\Phi}_1}{\partial \hat{z}}\right)\bigg|_{\hat{z}=1} \\ \left.\dfrac{\partial \hat{\Phi}_2}{\partial \hat{z}}\right|_{\hat{z}=0} = 0 \\ \hat{\Phi}_2|_{\hat{t}=0} = 0 \end{cases} \quad (C4)$$

Here we only list the first three sets of equations. The solution to Eq. (C2) has been obtained in Appendix A.1. Once $\hat{\Phi}_0$ is obtained, Eq. (C3) is a linear problem which can be solved easily by the separation of variable method. So is Eq. (C4) when $\hat{\Phi}_0$ and $\hat{\Phi}_1$ are obtained.

The expressions of $\hat{\Phi}_1$ is given as

$$\hat{\Phi}_1 = \frac{1}{2+Bi}\hat{z}^2 q_1(\hat{t}) + \sum_{n=1}^{\infty} T_{1n}(\hat{t})\cos(\omega_n \hat{z}) \tag{C5}$$

where

$$T_{1n}(\hat{t}) = \sum_{r=0}^{\infty}\sum_{s=1}^{\infty} C_{rsn} A_r A_s \frac{1}{\omega_r^2 + \omega_s^2 - \omega_n^2}\left[e^{-\omega_n^2 \hat{t}} - e^{-(\omega_r^2+\omega_s^2)\hat{t}}\right] + B_{1n} e^{-\omega_n^2 \hat{t}} \tag{C6}$$

$$q_1(t) = \sum_{r=1}^{\infty}\sum_{s=1}^{\infty} A_r A_s e^{-(\omega_r^2+\omega_s^2)\hat{t}} \omega_s \cos\omega_r \sin\omega_s + \sum_{s=1}^{\infty} A_s \omega_s e^{-\omega_s^2 \hat{t}} \sin\omega_s \tag{C7}$$

$$B_{1n} = -a q_1(0)\left(\frac{1}{\omega_n}\sin\omega_n + \frac{2}{\omega_n^2}\cos\omega_n - \frac{2}{\omega_n^3}\sin\omega_n\right) \tag{C8}$$

$$\begin{aligned}
C_{rsn} &= \frac{1}{4}\omega_s \omega_r\left[\frac{1}{\omega_s-\omega_r+\omega_n}\sin(\omega_s-\omega_r+\omega_n) + \frac{1}{\omega_s-\omega_r-\omega_n}\sin(\omega_s-\omega_r-\omega_n)\right] \\
&\quad - \frac{1}{4}\omega_s \omega_r\left[\frac{1}{\omega_s+\omega_r+\omega_n}\sin(\omega_s+\omega_r+\omega_n) + \frac{1}{\omega_s+\omega_r-\omega_n}\sin(\omega_s+\omega_r-\omega_n)\right] \\
&\quad - \frac{1}{4}\omega_s^2\left[\frac{1}{\omega_s+\omega_r+\omega_n}\sin(\omega_s+\omega_r+\omega_n) + \frac{1}{\omega_s+\omega_r-\omega_n}\sin(\omega_s+\omega_r-\omega_n)\right] \\
&\quad - \frac{1}{4}\omega_s^2\left[\frac{1}{\omega_s-\omega_r+\omega_n}\sin(\omega_s-\omega_r+\omega_n) + \frac{1}{\omega_s-\omega_r-\omega_n}\sin(\omega_s-\omega_r-\omega_n)\right] \\
&\quad + \frac{2}{2+Bi}\frac{\omega_s}{\omega_n}\cos\omega_r \sin\omega_s \sin\omega_n \\
&\quad + \frac{1}{2+Bi}(\omega_r^2+\omega_s^2)\omega_s \cos\omega_r \sin\omega_s\left(\frac{1}{\omega_n}\sin\omega_n + \frac{2}{\omega_n^2}\cos\omega_n - \frac{2}{\omega_n^3}\sin\omega_n\right)
\end{aligned} \tag{C9}$$

while $A_n$ and $\omega_n$ are the same as in Appendix A.1.

The expressions of $\hat{\Phi}_2$ is given as

$$\hat{\Phi}_2(\hat{z},\hat{t}) = \frac{1}{2+Bi}\hat{z}^2 q_2(\hat{t}) + \sum_{n=1}^{\infty} T_{2n}(\hat{t})\cos(\omega_n \hat{z}) \tag{C10}$$

where

$$T_{2n}(\hat{t}) = B_{2n} e^{-\omega_n^2 \hat{t}} + \int_0^{\hat{t}} e^{-\omega_n^2(\hat{t}-\tau)} r_{2n}(\tau)\,d\tau \tag{C11}$$

$$q_2(\hat{t}) = \left( \frac{1}{2+Bi} q_1(\hat{t}) \sum_{r=1}^{\infty} T_{0r}(\hat{t}) \omega_r \sin \omega_r + \sum_{r=1}^{\infty} \sum_{s=1}^{\infty} T_{1s}(\hat{t}) T_{0r}(\hat{t}) \omega_r \sin \omega_r \cos \omega_s \right)$$

$$- \left( \frac{2}{2+Bi} q_1(\hat{t}) + \frac{2}{2+Bi} q_1(\hat{t}) \sum_{r=1}^{\infty} T_{0r}(\hat{t}) \cos \omega_r - \sum_{s=1}^{\infty} T_{1s}(\hat{t}) \omega_s \sin \omega_s \right)$$

$$+ \left( \sum_{r=1}^{\infty} \sum_{s=1}^{\infty} T_{1s}(\hat{t}) T_{0r}(\hat{t}) \omega_s \cos \omega_r \sin \omega_s \right)$$

(C12)

$$B_{2n} = -\frac{q_2(0)}{2+Bi} \left( \frac{1}{\omega_n} \sin \omega_n + \frac{2}{\omega_n^2} \cos \omega_n - \frac{2}{\omega_n^3} \sin \omega_n \right) \tag{C13}$$

$$r_{2n}(\hat{t}) = p_{2n}(\hat{t}) + \frac{2}{2+Bi} q_2(\hat{t}) \frac{1}{\omega_n} \sin \omega_n - a q_2'(\hat{t}) \left( \frac{1}{\omega_n} \sin \omega_n + \frac{2}{\omega_n^2} \cos \omega_n - \frac{2}{\omega_n^3} \sin \omega_n \right)$$

(C14)

$$p_{2n}(\hat{t}) = -4a q_1(\hat{t}) \sum_{r=1}^{\infty} T_{0r}(\hat{t}) \omega_r \left\{ \begin{array}{l} \frac{1}{2} \frac{1}{\omega_r + \omega_n} \left[ -\cos(\omega_r + \omega_n) + \frac{1}{\omega_r + \omega_n} \sin(\omega_r + \omega_n) \right] + \\ \frac{1}{2} \frac{1}{\omega_r - \omega_n} \left[ -\cos(\omega_r - \omega_n) + \frac{1}{\omega_r - \omega_n} \sin(\omega_r - \omega_n) \right] \end{array} \right\}$$

$$-a q_1(\hat{t}) \sum_{r=1}^{\infty} T_{0r}(\hat{t}) \omega_r^2 \left\{ \begin{array}{l} \frac{1}{2} \frac{1}{\omega_r + \omega_n} \left[ \sin(\omega_r + \omega_n) + \frac{2}{\omega_r + \omega_n} \left( \cos(\omega_r + \omega_n) - \frac{1}{\omega_r + \omega_n} \sin(\omega_r + \omega_n) \right) \right] + \\ \frac{1}{2} \frac{1}{\omega_r - \omega_n} \left[ \sin(\omega_r - \omega_n) + \frac{2}{\omega_r - \omega_n} \left( \cos(\omega_r - \omega_n) - \frac{1}{\omega_r - \omega_n} \sin(\omega_r - \omega_n) \right) \right] \end{array} \right\}$$

$$-2 \sum_{r=1}^{\infty} \sum_{s=1}^{\infty} T_{1s}(\hat{t}) T_{0r}(\hat{t}) \omega_r^2 \frac{1}{4} \left[ \begin{array}{l} \frac{1}{\omega_r + \omega_s + \omega_n} \sin(\omega_r + \omega_s + \omega_n) + \frac{1}{\omega_r + \omega_s - \omega_n} \sin(\omega_r + \omega_s - \omega_n) \\ + \frac{1}{\omega_r - \omega_s + \omega_n} \sin(\omega_r - \omega_s + \omega_n) + \frac{1}{\omega_r - \omega_s - \omega_n} \sin(\omega_r - \omega_s - \omega_n) \end{array} \right]$$

$$+2 \sum_{r=1}^{\infty} \sum_{s=1}^{\infty} T_{1s}(\hat{t}) T_{0r}(\hat{t}) \omega_r \omega_s \frac{1}{4} \left[ \begin{array}{l} \frac{1}{\omega_r - \omega_s + \omega_n} \sin(\omega_r - \omega_s + \omega_n) + \frac{1}{\omega_r - \omega_s - \omega_n} \sin(\omega_r - \omega_s - \omega_n) \\ - \frac{1}{\omega_r + \omega_s + \omega_n} \sin(\omega_r + \omega_s + \omega_n) - \frac{1}{\omega_r + \omega_s - \omega_n} \sin(\omega_r + \omega_s - \omega_n) \end{array} \right]$$

$$+2a q_1(\hat{t}) \frac{1}{\omega_n} \sin \omega_n + 2a q_1(\hat{t}) \sum_{r=1}^{\infty} T_{0r}(\hat{t}) \frac{1}{2} \left[ \frac{1}{\omega_r + \omega_n} \sin(\omega_r + \omega_n) + \frac{1}{\omega_r - \omega_n} \sin(\omega_r - \omega_n) \right]$$

$$- \sum_{s=1}^{\infty} T_{1s}(\hat{t}) \omega_s^2 \frac{1}{2} \left[ \frac{1}{\omega_s + \omega_n} \sin(\omega_s + \omega_n) + \frac{1}{\omega_s - \omega_n} \sin(\omega_s - \omega_n) \right]$$

(C15)

$$a = \frac{1}{2+Bi} \tag{C16}$$

As a numerical example, we calculate the maximum DIS in contraction diffusion using the three-term solution (second order with respect to $\hat{\beta}$). As shown by Fig C1(a), the perturbation solution $\hat{\Phi}(\hat{z},\hat{t}) = \hat{\Phi}_0(\hat{z},\hat{t}) + \hat{\beta}\hat{\Phi}_1(\hat{z},\hat{t}) + \hat{\beta}^2\hat{\Phi}_2(\hat{z},\hat{t})$ is almost accurate when $\hat{\beta}=1$ for various Biot numbers. The accurate solution here is obtained by finite difference method using the code 'pdepe' in the commercial software Matlab (Version 2012a). Fig C1 (b)(c)(d) show the regions where the perturbation solution has a relative error less than 10% in the $Bi - \hat{\beta}$ plane. We can see that the perturbation solution fails when $\hat{\beta}>1$. It should be pointed out that for $\hat{\beta}=0$, i.e. the stress-independent-diffusion case, our half-analytical solution in series form is in complete agreement with the results by the finite difference method, implying the validation of both approaches.

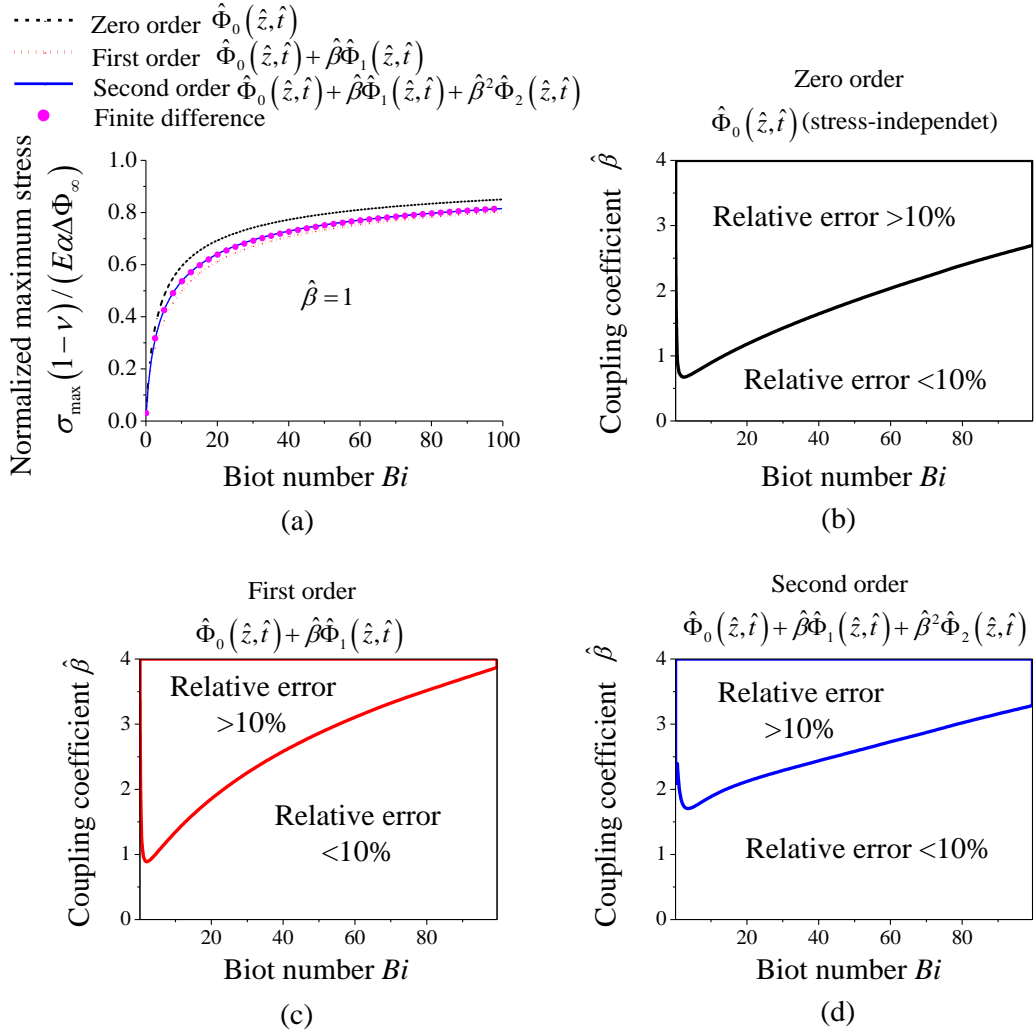

**Fig C1.** (a)The normalized maximum DIS calculated by the perturbation solution when the coupling coefficient $\hat{\beta}=1$. (b)(c)(d) Illustration of the region where the maximum diffusion induced stress predicted by the perturbation solution gives rise to a relative error less than 10%.